\documentclass[twocolumn]{aastex62}

\begin{document}

\title{Dusty Superwind from a Galaxy with a Compact Obscured Nucleus: Optical Spectroscopic Study of NGC~4418}

\shorttitle{Dusty Superwind from NGC~4418}

\author{Youichi Ohyama}
\author{Kazushi Sakamoto}
\affiliation{Academia Sinica, Institute of Astronomy and Astrophysics, 11F of Astronomy-Mathematics Building, AS/NTU, No.1, Sec. 4, Roosevelt Rd, Taipei 10617, Taiwan, R.O.C.}

\author{Susanne Aalto}
\affiliation{Department of Earth and Space Sciences, Chalmers University of Technology, Onsala Space Observatory, SE-439 92, Onsala, Sweden}

\author{John S. Gallagher III}
\affiliation{Department of Astronomy, University of Wisconsin-Madison, 5534 Sterling, 475 North Charter Street, Madison, WI 53706}

\shortauthors{Ohyama et al.}

\begin{abstract}
We report our optical spectroscopic study of the nucleus and its surrounding region of a nearby luminous infrared galaxy NGC~4418.
This galaxy has been known to host a compact obscured nucleus, showing distinct characteristics such as a very compact ($\sim 20$~pc) sub-mm and mid-infrared core and dusty circumnuclear region with massive molecular gas concentration.
We detected dusty superwind outflow at $\gtrsim 1$~kpc scale along the disk semiminor axis in both shock-heated emission lines and enhanced interstellar Na~D absorption.
This superwind shows basic characteristics similar to those of the prototypical superwind in the starburst galaxy M82, such as a kpc-scale extended structure of gas and dust along the disk minor axis, outflowing components (multiphase gas and dust), physical conditions of the ionized gas, and monotonically blueshifting radial velocity field with increasing distance from the nucleus on the front side of the superwind.
We also detected a moderately extinct starburst population in the SDSS nuclear spectrum with the burst age of $\simeq 10$~Myr and stellar mass of $\simeq 1\times 10^7\ M_\mathrm{\sun}$.
It is powerful enough to drive the superwind within the dynamical age of the superwind ($\simeq 10$~Myr).
On the basis of comparison between this starburst--superwind scenario and the observations in terms of the burst age, stellar mass, infrared luminosity, and obscuration in the optical bands, we argue that this superwind-driving starburst is separate from the sub-mm core even if the core is a very young star cluster.
Therefore, this galaxy hosts both the enshrouded compact core and the superwind-driving circumnuclear starburst.
\end{abstract}


\section{INTRODUCTION}\label{intro}

There is a class of galaxies that harbor very compact (typically smaller than several tens pc) infrared-luminous nuclei.
Such nuclei are often highly obscured at optical and near-infrared (NIR) wavelengths ($A_\mathrm{V}\gtrsim 100$ mag), and are referred to as compact obscured nuclei (CONs; e.g., \citealt{costagliola11}).
The best known examples are IC~860 (e.g., \citealt{costagliola11}), NGC~1377 (e.g., \citealt{roussel06,aalto12}), and NGC~4418 (e.g., \citealt{costagliola13,sakamoto13}).
They are peculiar at far infrared (FIR) and radio wavelengths.
They show large FIR excess over the radio compared to regular star-forming galaxies.
This is surprising because both FIR and radio luminosities are known to tightly correlate with star forming activities and, therefore, their luminosity ratios are within a tight range for star-forming galaxies (e.g., \citealt{yun01,roussel03,roussel06}).
The radio spectral indices of CONs are very flat (e.g., \citealt{condon91b,roussel03,rg07}), suggesting either that thermal radio emission dominates or that synchrotron self-absorption or free-free absorption is significant.
A clear trend is known that galaxies with larger FIR-to-radio ratios have hotter FIR color temperatures as measured in {\it IRAS} $60\ \mu$m and $100\ \mu$m fluxes ($S_\mathrm{60\mu m}/S_\mathrm{100\mu m}$; e.g., \citealt{yun01,roussel03}).
Such sources are very rare, and \cite{yun01} therefore argued that their lifetime should be very short (only about 1 percent of an infrared bright phase; see also \citealt{roussel03}).

An idea of nascent starburst has been proposed to explain the peculiar radio and FIR properties of CONs (e.g., \citealt{yun01,bressan02,roussel03,rg07}).
In this scenario, the starburst within CONs is younger than the main-sequence lifetime of most massive stars (4--6~Myr) and, hence, has little non-thermal radio emission due to supernovae (SNe).
The UV radiation density in and around such compact young starburst is expected to be high, being consistent with the higher dust temperature and physical conditions of the molecular gas revealed by the molecular line analysis (e.g., \citealt{roussel06,aalto07,costagliola10,costagliola11,aalto12,costagliola13,sakamoto13}).
Nascent starbursts are rare by definition because they are in the first few Myr of starbursts and they will soon become ordinary starbursts with many SNe.
The starburst nucleus is expected to blow out much of its dusty shroud in this process.
If CONs are nascent starbursts, then we probably witness in them beginning/onset of this transition process.
An idea of hidden active galactic nucleus (AGN) has been also advocated for the main luminosity source of CONs.
If a radio-quiet AGN is buried deeply enough in dust and gas, then it can also have a compact and hot core that radiates predominantly in FIR, without showing usual optical and X-ray signatures of an AGN.
In this scenario, it seems also likely that the dusty cocoon around the AGN is soon expelled due to strong AGN feedback (radiation pressure and/or relativistic jet).
The distinction between compact starburst and AGN for the energy source of CONs has been a matter of debate.
In either case, such objects likely provide us with chances to investigate the feedback process on the circumnuclear material around the compact nucleus.
In addition to the studies of the population/powering source of CONs, we also want to know how and why such very compact dusty structure is formed in some nuclei.
This is because some well known (ultra) luminous infrared galaxies ((U)LIRGs) also harbor similar very compact nuclei from which most of the energy is released (e.g., the western nucleus of Arp~220; e.g., \citealt{sakamoto08,sakamoto17}).

NGC~4418 is one of the closest and best studied galaxies hosting CONs.
It is a LIRG with its 8--$1000\ \mu$m infrared luminosity ($L_\mathrm{IR}$) of $1.4\times 10^{11}\ L_\mathrm{\sun}$ \citep{sanders03}\footnote{With our assumed distance, the luminosity is larger than the threshold for the LIRG classification of $L_\mathrm{IR}=1\times 10^{11}\ L_\mathrm{\sun}$ \citep{sanders96}. We note, however, that NGC~4418 is sometimes classified as a non-LIRG if one adopts smaller distance to the galaxy. For example, with its recession velocity ($cz$) and an assumed Hubble constant of $H_\mathrm{0}=75$ km~s$^{-1}$~Mpc$^{-1}$, the distance is estimated to be 28.32~Mpc and $L_\mathrm{IR}=9.4\times 10^{10}\ L_\mathrm{\sun}$.} at a distance of 34~Mpc ($1''=165$~pc; \citealt{sakamoto13})\footnote{\cite{sakamoto13} adopted the distance from \cite{sanders03}, in which distance is calculated with $cz$ using the cosmic attractor model outlined in Appendix A of \cite{mould00}, using $H_\mathrm{0}=75$ km~s$^{-1}$~Mpc$^{-1}$ and adopting a flat cosmology in which $\Omega_\mathrm{M}=0.3$ and $\Omega_\lambda=0.7$. We adopt this distance for consistency with recent closely-related studies of this galaxy \citep{aalto12,costagliola13,varenius14}.}.
In spite of its huge infrared luminosity, the host galaxy is classified from the optical morphology as an ordinary Sa-type galaxy ((R')SAB(s)a; \citealt{rc3}) with moderate inclination ($i=62^\circ$; taken from \citealt{sakamoto13} for consistency; based on \citealt{2mass}).
It has a neighbor galaxy about 32~kpc ($3\farcm 2$) away (VV~655=MCG+00-32-013=SDSS J122704.47-005420.6; $cz_\mathrm{SDSS;~helio.}=2204$ km~s$^{-1}$; \citealt{evans03,varenius17}).
\cite{varenius17} found a bridge of H~{\sc i} 21~cm emission connecting these two galaxies and argued their strong tidal interaction about 190~Myr ago.
The optical nuclear emission-line spectrum of NGC~4418 is classified as LINER \citep{armus89,shi05} or Seyfert 2 \citep{baan98}, but no direct signature of AGN such as very broad permitted lines is known (see also \citealt{lehnert95}).
NGC~4418 is deficient of X-ray emission for its infrared luminosity, and its interpretation has been in controversy.
\cite{maiolino03} tentatively identified, with low statistics, a Compton-thick (reflection-dominated) AGN on the basis of its small photon index (or flat X-ray spectrum; $\Gamma=0.76^{+0.61}_{-0.62}$, where $N(E)\propto E^{-\Gamma}$, $E$ is the photon energy, and $N(E)$ is the photon number density).
On the other hand, \cite{lehmer10} found the opposite (soft) spectral index ($\Gamma=3.28\pm 0.46$) that is even softer than the one of Compton-thin AGNs ($\Gamma\simeq 1.7$; \citealt{mushotzky93}).
We note that both analyses are based on the {\it Chandra} AICS-S data, and \cite{lehmer10} added $\simeq 30$\% more integration onto the earlier data used by \cite{maiolino03}.

The FIR and radio properties of NGC~4418 are distinct from those of ordinary LIRG(s).
Its FIR color temperature ($S_\mathrm{60\mu m}/S_\mathrm{100\mu m}>1$; \citealt{sanders03}), or $T_\mathrm{d}=60$--80~K (for spectral dust emissivity index $\beta=1$, where $Q_\lambda\propto \lambda^{-\beta}$), is one of the largest among normal star-forming galaxies (e.g., \citealt{dale00,chapman03}).
The FIR luminosity is $\sim 5$ times larger than that of typical galaxies with similar radio luminosity \citep{yun01,condon02,roussel03}.
The FIR-to-radio luminosity ratio is often parameterized as $q\equiv\log((\mathrm{FIR}/3.75\times 10^{12}$~W m$^{-2})/(S_\mathrm{1.4~GHz}$~W m$^{-2}$ Hz$^{-1}$)), where FIR$\equiv 1.26\times 10^{-24}\times(2.58\times S_\mathrm{60\mu m}+S_\mathrm{100\mu m})$ (\citealt{helou85,condon91}; see also \citealt{condon92}).
NGC~4418 shows $q\simeq 3.1$, which is $>3$ sigma above the average ratio of $q=2.34$ \citep{condon02,roussel03}.
It is one of the only nine galaxies showing $q>3$ in the {\it IRAS} 2~Jy sample containing 1809 galaxies (\citealt{yun01}; see also \citealt{roussel03} for the revised statistics).
The radio emission is confined within the central $\lesssim 0\farcs 5$ (\citealt{condon90,baan06,costagliola13,varenius14}. See also Section~\ref{consistency_radio}).
The radio spectral index ($\alpha$, where $S_\nu\propto\nu^\alpha$) between 4.9 and 1.4~GHz is measured, at the scale of a few arcsec or larger, to be $\alpha=-0.26$ \citep{carilli00} and $\alpha=-0.30$ \citep{roussel03}.
This is significantly flatter than the typical synchrotron spectrum of spiral galaxies ($\alpha\simeq -0.7$; \citealt{condon91b}).
The flat radio spectrum continues down to 0.33~GHz with $\alpha=-0.27$ between 1.4~GHz and 0.33~GHz \citep{carilli00}.

A very compact core of NGC~4418 has been studied at both mid-infrared (MIR) and sub-mm wavelengths.
The MIR core is very compact and almost unresolved with the Keck telescope at diffraction-limited condition (at full-width half maximum (FWHM) resolutions of $0\farcs 23$ (30~pc) at $8\ \mu$m, $0\farcs 31$ (40~pc; marginally resolved) at $10\ \mu$m, and $0\farcs 61$ (80~pc) at $24.5\ \mu$m; \citealt{evans03}) and VLT ($<50$~pc at $11.25\ \mu$m; \citealt{siebenmorgen08}).
The MIR surface brightness of the nucleus is much higher than that of typical nearby starburst galaxies (\citealt{siebenmorgen08}; see also \citealt{evans03} for discussion on the infrared (8--$1000\ \mu$m) surface brightness).
The nucleus shows a very deep silicate absorption feature at $9.7\ \mu$m, indicating huge absorption corresponding to $A_\mathrm{V}\sim 50$--130 mag toward the MIR core \citep{roche86,dudley97,spoon01,evans03,siebenmorgen08,roche15}.
The enhanced MIR emission of the core is visible down to $\sim 5\ \mu$m \citep{spoon01,evans03,siebenmorgen08}, but the corresponding core at $2\ \mu$m is not in the {\it HST} image \citep{scoville00,evans03}.
At sub-mm wavelengths, meanwhile, a very compact ($\sim 20$~pc) core is found (hereafter, the sub-mm core; \citealt{sakamoto13}).
This core is surrounded by a massive molecular gas concentration ($\sim 10^{8.0}\ M_\mathrm{\sun}$ at $r\lesssim 10$~pc; \citealt{costagliola13,sakamoto13}), and the dynamical mass measured with the gas kinematics is $\sim 10^{8.3\pm 0.3}\ M_\mathrm{\sun}$ at $r\lesssim 15$~pc; \citealt{sakamoto13}.
\cite{sakamoto13} constrained physical properties of the sub-mm core, and showed that a very compact young ($\lesssim 5$~Myr) starburst with stellar mass of $\sim 10^{8.0\pm 0.5}\ M_\mathrm{\sun}$ at $r\lesssim 10$~pc as well as an AGN can reproduce the observations.
\cite{varenius14} recently conducted VLBI observation at cm wavelengths, and resolved the sub-mm core into eight very compact ($<8$~pc) sources within the central 41~pc.
They argued that a single AGN cannot dominate the core and many of the individual blobs, if not all, are likely very compact star clusters.

In this paper, we study a large-scale nebula of both ionized and cold gases of NGC~4418 via optical spectroscopy, and discuss its nature and its driving source at the nucleus.
The presence of an extended ionized-gas nebula at a kpc scale has been reported by \cite{lehnert95}.
A large-scale superwind was first proposed by \cite{sakamoto13}, who identified a kpc-scale extinct region on an optical color map.
The extinction is strongest near the nucleus, and extends preferentially along a semiminor axis of the disk, forming the shape of a letter ``U''.
{\it HST} NIR images show moderately extinct nucleus ($A_\mathrm{V}\sim 2$--5 mag) behind radial dust lanes at the scale of 100--200~pc \citep{scoville00,evans03}, and such features might be connected to the kpc-scale extinct region.
\cite{sakamoto13} argued that such region traces a galaxy-scale outflow of dusty material, although little kinematical evidence was available.
Recently, \cite{varenius17} detected blueshifted H~{\sc i} emission at $\simeq 1'$ ($\simeq 10$~kpc) north-west (hereafter, NW) of the nucleus, and argued that the nuclear outflow proposed by \cite{sakamoto13} is responsible for it.
We note that the H~{\sc i} emission was detected well outside the optical size of this galaxy ($27''$ in the $r'$-band Petrosian radius; SDSS; see, e.g., \citealt{yasuda01} for definition of the radius) and at far away from the region where \cite{sakamoto13} examined the dust reddening (up to $\simeq 10''$).
We aim to reveal the detailed characteristics of the superwind to constrain the superwind-driving activity at the nucleus by applying analyses often used for typical starbursts with superwind activities (e.g., \citealt{heckman90,heckman00,veilleux05}).
We also aim to infer the obscured activity of the core.

\section{OBSERVATIONS AND DATA}

\subsection{Optical Broad-Band Imaging}\label{imag_obs}

\begin{figure*}
\begin{center}
\includegraphics[angle=-90,scale=0.65]{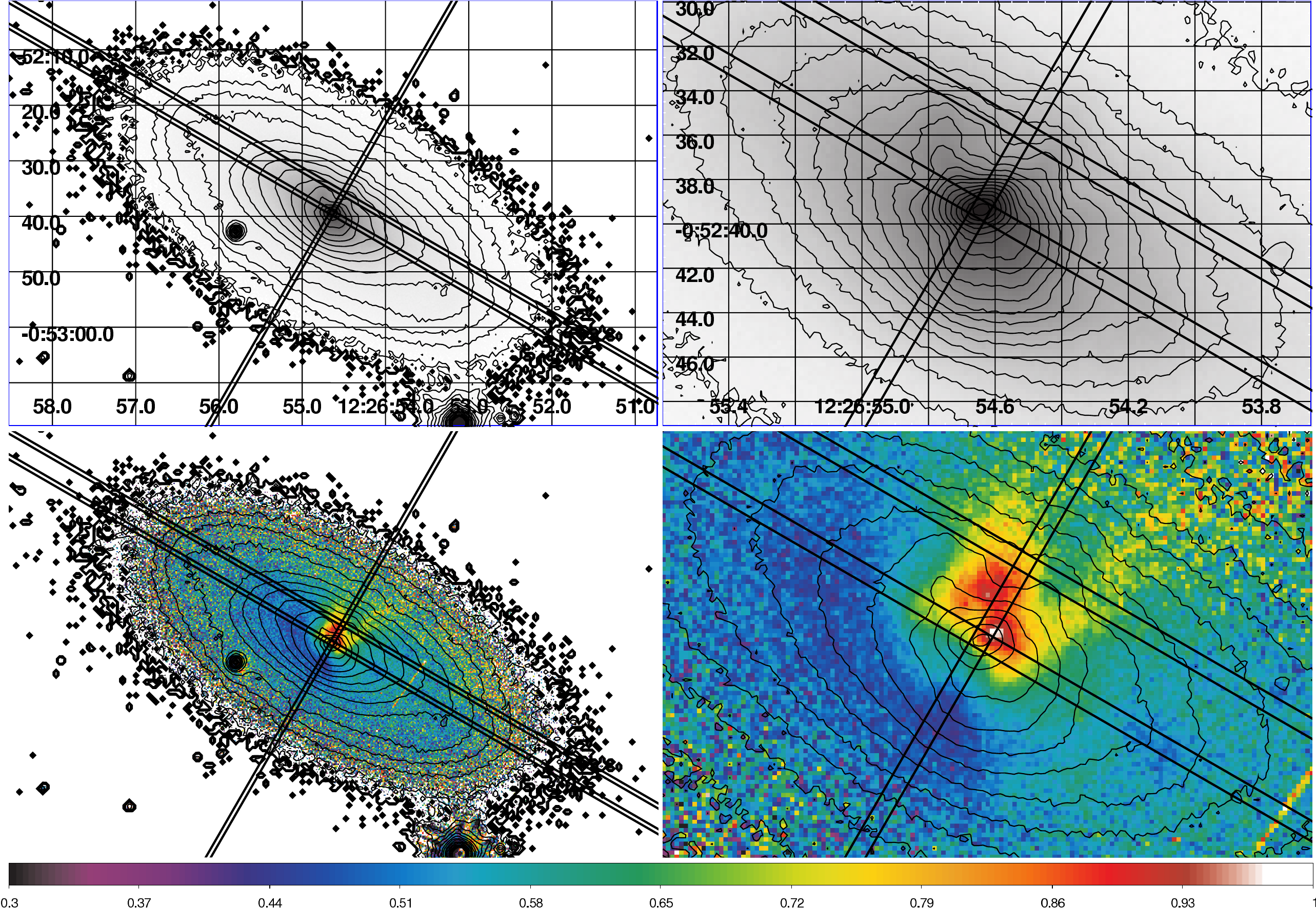}
\caption{
The KIDS $g'$- (top) and $g'-r'$ (AB) color (bottom) images of NGC~4418 in ASNIH scale.
North is up and east is to the left.
The color coding of the $g'-r'$ image is shown in the bottom color bar.
Right column images zoom up the nuclear region of the corresponding images in the left column.
Region with lower S/N ($<3$) is masked (white).
Three slit locations of our spectroscopy are shown with pairs of parallel lines.
The $g'$-band image contours are drawn in log steps for positional reference in all panels.
\label{findingchart}}
\end{center}
\end{figure*}

NGC~4418 was observed as a part of KIDS survey \citep{kids} in optical bands, and we retrieved fully processed images from its DR2.
Galactic extinction ($A_\mathrm{V}=0.064$ mag; NED) is corrected by assuming an extinction law of \cite{ccm} with $R=3.1$.
Figure~\ref{findingchart} shows the $g'$-band and $g'-r'$ (AB) images.
Seeing size is $0\farcs 9$--$1\farcs 0$ FWHM in both bands.
They are deeper and higher in spatial resolution than the SDSS color image presented by \cite{sakamoto13}.

\subsection{Optical Spectroscopy}

\subsubsection{FOCAS/Subaru observations and data}\label{spec_obs}

The long-slit spectra of NGC~4418 were taken with FOCAS spectrograph \citep{focas} at the Subaru telescope on March 8, 2013.
We placed a 0\farcs 8-wide slit at three locations (Figure~\ref{findingchart}):
along the disk minor axis (position angel (PA)$=150^\circ$), the disk major axis (PA$=60^\circ$), and parallel to the disk major axis (PA$=60^\circ$) but at $4''$ off the nucleus toward NW (hereafter, the offset-major axis slit).
We adopted the major-axis PA of the galaxy disk (PA$=60^\circ$) from the 2MASS extended source catalog \citep{2mass}.
Seeing was variable during the observations, ranging 0\farcs 5 (for the minor axis slit)--1\farcs 8 (for both major and offset-major axis slits) FWHM as measured on the target acquisition images taken with the FOCAS imaging mode just before starting the spectroscopy exposures.
The CCD pixels were binned on-chip to $2\times 2$ pixels ($0\farcs 2\times 0\farcs 2$) before reading out.
We used an Echelle grism with an order sorting filter of SDSS $r'$ to isolate the third order spectrum at 5500\AA--6900\AA~to cover Na~D $\lambda\lambda$5890, 5896 (Na~D2 and D1, respectively), stellar photospheric absorption at 6494\AA~(see below), H$\alpha$, [N~{\sc ii}]$\lambda\lambda$6548, 6583, and [S~{\sc ii}]$\lambda\lambda$6716, 6731.
The stellar absorption at 6494\AA, which is often seen in spectra of star clusters and galaxies (e.g., \citealt{heckman95,westmoquette07}), is a blend of Ca~{\sc i} and Fe~{\sc i} lines (hereafter, Ca~{\sc i}+Fe~{\sc i}) originating mostly from late-type stars (e.g., \citealt{mora01}).
We obtained three 10-minute exposures per slit for stacking.
The data reduction was made in a standard way for long-slit spectroscopy, including overscan subtraction of the CCD, flat fielding by using the dome flat exposures, wavelength calibration by using sky OH lines, and spectral response calibration by using standard-star observations.
The RMS uncertainty of the wavelength calibration is typically 10 km~s$^{-1}$ over the entire two-dimensional spectral images.
The instrumental line width measured with the OH lines is 4.1\AA~FWHM at all wavelengths, giving the spectral resolution ($R \equiv \lambda/\Delta \lambda$) of 1600 and 1440 around H$\alpha$ and Na~D, respectively.
We show the recession velocities in the Local Standard of Rest frame throughout this paper, if not explicitly mentioned.

\subsubsection{Spectral fitting}\label{spectral_fitting}

\begin{figure*}
\plotone{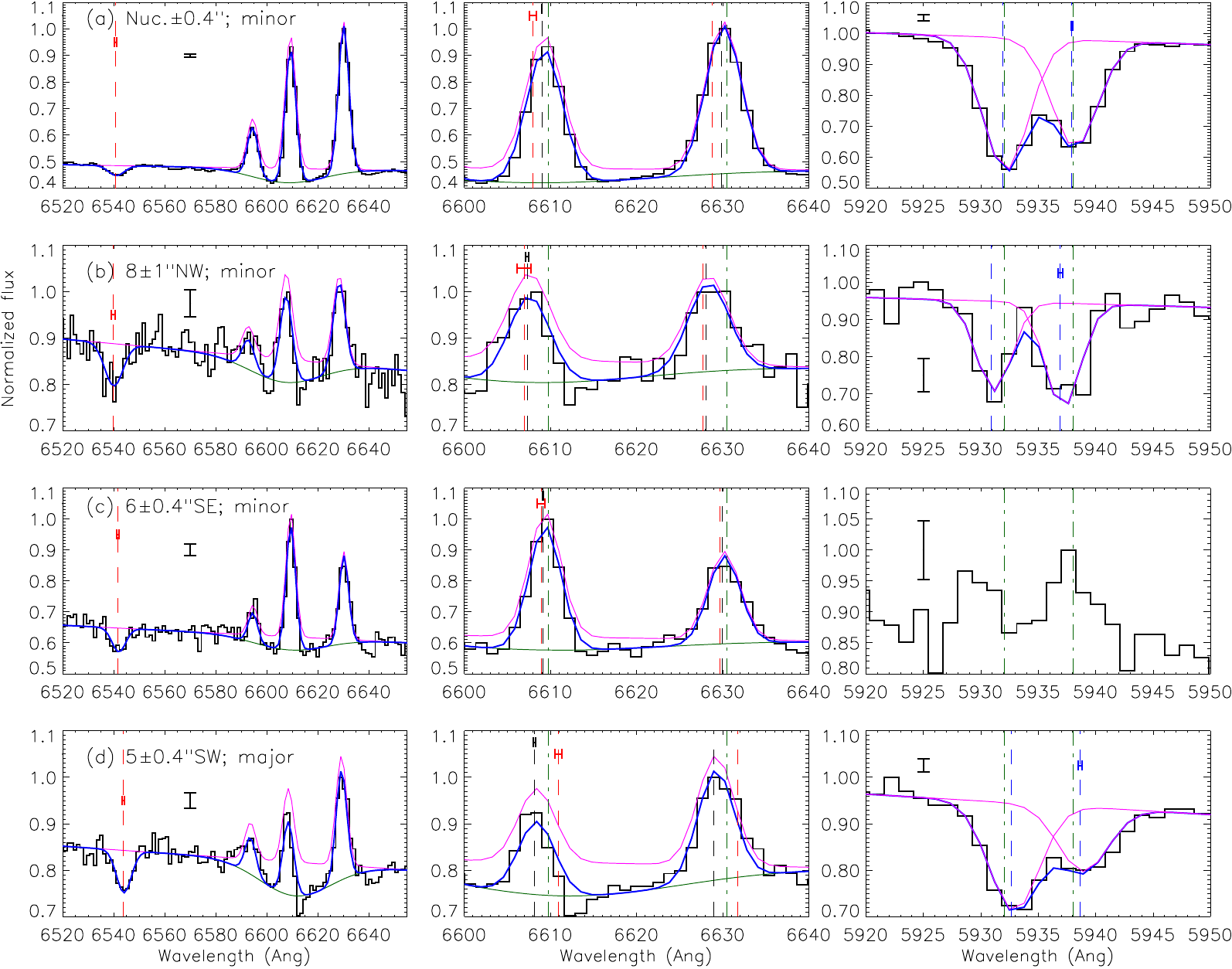}
\caption{The normalized FOCAS spectra and the fit results at selected regions of interests.
Left, central, and right panels show H$\alpha$ and [N~{\sc ii}] features with Ca~{\sc i}+Fe~{\sc i} absorption, only H$\alpha$ and [N~{\sc ii}]$\lambda 6583$ features, and Na~D doublet features, respectively.
Typical one-sigma uncertainties of the spectra are indicated with vertical error bars in both left and right panels.
The information of the four selected regions (central pixel positions and width for integrating the spectra) are indicated in the left panels:
(a) The nucleus; (b) near the NW tip of the nebula along the semiminor axis; (c) the opposite side of position (b) along the semiminor axis; and (d) the SW side of the nucleus on the disk along the major axis.
See text for more descriptions about these regions.
Our best fit profiles are shown for H$\alpha$+[N~{\sc ii}] emission lines on the stellar continuum (magenta), Na~D1 and D2 absorption lines on the stellar continuum (magenta), and Ca~{\sc i}+Fe~{\sc i} and H$\alpha$ absorption lines on the stellar continuum (green).
The total fit spectra (sum of the stellar continuum, stellar absorption, and emission/absorption lines of the ionized/cold gas) are shown in blue.
Long vertical lines indicate the best-fit wavelengths of Ca~{\sc i}+Fe~{\sc i} (red dashed), H$\alpha$+[N~{\sc ii}] (black dashed), and Na~D (blue dashed).
Wavelengths corresponding to the velocity of origin of the outflow are also shown in green dot-dashed vertical lines (see text for the details).
Uncertainties (one sigma) of these wavelengths are shown as horizontal error bars with the corresponding colors.
We note that the error bars are sometimes very tiny to clearly see in this figure.
The Na~D spectrum in panel (c), where Na~D is probably in emission, could not be fit.
\label{focas_profmap}}
\end{figure*}

We performed multi-component spectral fitting on both emission and absorption lines to extract information of ionized gas, cold gas, and stars.
The ionized-gas component consists of H$\alpha$, [N~{\sc ii}] doublet, and [S~{\sc ii}] doublet emission lines.
The [N~{\sc ii}] doublet ratio ([N~{\sc ii}]$\lambda 6583$/[N~{\sc ii}]$\lambda 6548$) was fixed to the theoretical value (3.0) during the fit.
We assumed that all emission lines share the same velocity field (i.e., we fit the emission lines with only one set of redshift and line width) at each position along the slits.
The cold-gas component consists of Na~D doublet absorption.
The stellar component consists of Ca~{\sc i}+Fe~{\sc i} and H$\alpha$ absorption lines on the stellar continuum.
We assumed different line widths for Ca~{\sc i}+Fe~{\sc i} and H$\alpha$ absorption lines, but assumed that they share the same redshift.
Each of the three components has different redshift and line width.
We adopted Gaussian function to represent all emission and absorption line profiles, and linear function to represent the stellar continuum.
We binned spatial pixels during the fit by 2 pixels ($0\farcs 4$) to match the seeing size to examine the spatial profile of the fit results, or larger when higher signal-to-noise ratio (S/N) is required (Figure~\ref{focas_profmap}).
At each binned pixel position, we separately fit blue ($\lambda\lambda$5872--6001\AA) and red ($\lambda\lambda$6480--6827\AA) wavelength ranges of the spectrum, because the stellar continuum cannot be fit with a single linear function over the entire wavelength range.
The spectral features in all different components were simultaneously fit within each spectral range.
In the red range, H$\alpha$ absorption is generally weaker than the H$\alpha$ emission.
When the fit with both emission and absorption lines gave poor fit on the absorption line, we re-fit the spectrum without the absorption.
Figure~\ref{focas_profmap} shows the spectra and the fit results at selected regions of interests for our later analysis.
Those regions include the nucleus (Figure~\ref{focas_profmap}a), tip of the ionized-gas nebula along the disk semiminor axis (Figure~\ref{focas_profmap}b), the region showing the local excess of H$\alpha$ flux (Figure~\ref{focas_profmap}c), and the disk region where the ionized-gas velocity field is decoupled from those of the neutral gas and stellar component (Figure~\ref{focas_profmap}d; see Section~\ref{results_spatial_distribution} for the detailed descriptions of these regions).

\begin{figure*}
\epsscale{0.9}
\plotone{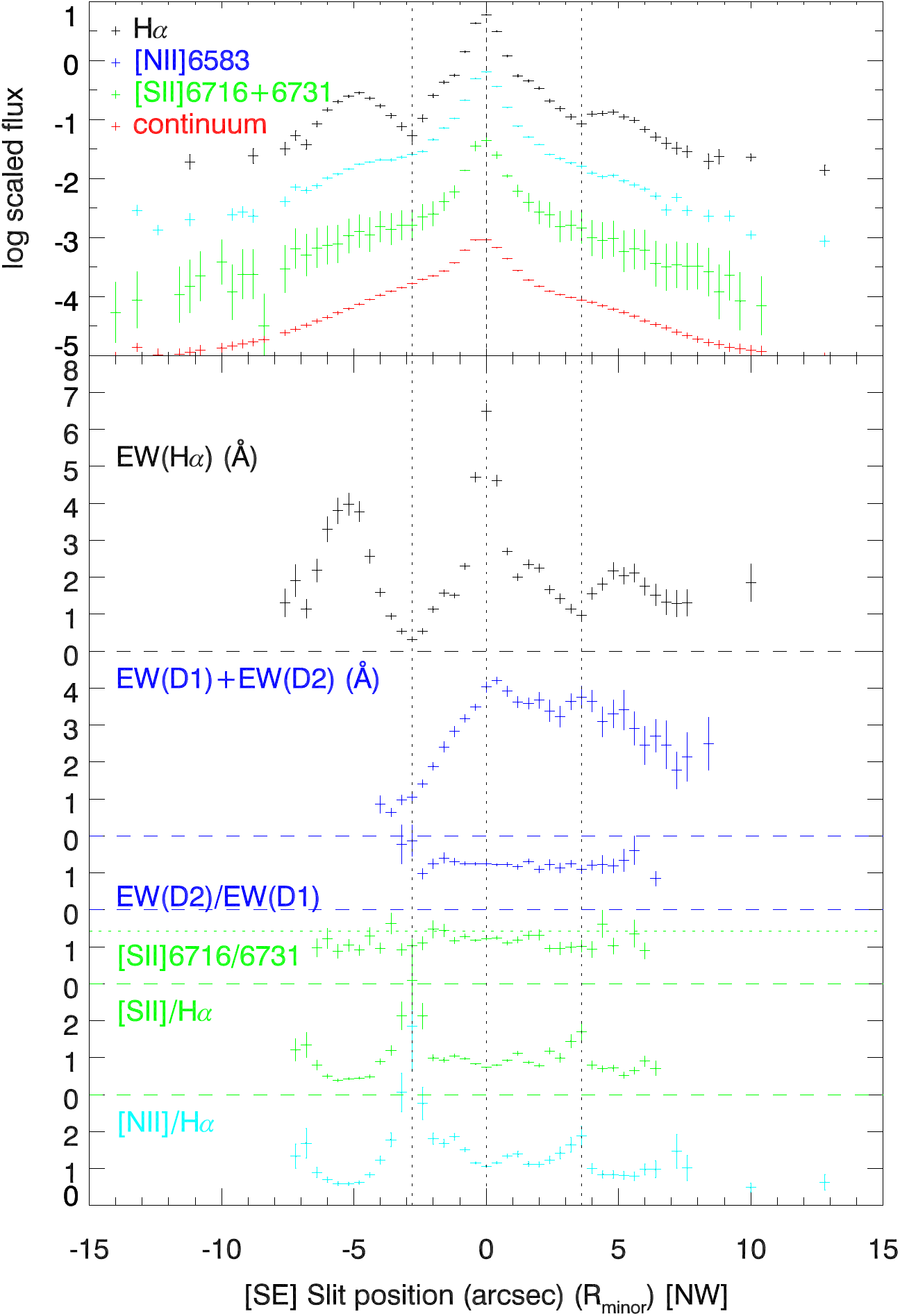}
\caption{Spatial distributions of fluxes, EWs, EW ratios, and line ratios along the disk minor axis.
(Top) fluxes of H$\alpha$ emission (black), [N~{\sc ii}]$\lambda 6583$ (cyan), [S~{\sc ii}]$\lambda 6716+\lambda 6731$ (green), and stellar continuum at H$\alpha$ (red) in log scale with arbitrary offsets for clarity of the figure.
(Bottom) EWs of H$\alpha$ emission (\AA; black) and Na~D absorption ($EW_\mathrm{Na~D1}+EW_\mathrm{Na~D2}$; \AA; blue), Na~D doublet ratio ($EW_\mathrm{Na~D2}/EW_\mathrm{Na~D1}$; blue), [S~{\sc ii}] doublet ratio (green), [S~{\sc ii}]$/$H$\alpha$ (green), and [N~{\sc ii}]$/$H$\alpha$ (cyan).
Zero levels of each sub-panel are shown with horizontal broken lines with corresponding colors.
In a sub-panel of [S~{\sc ii}] doublet ratio, the ratio for the low-density limit (1.43) is marked with a green horizontal dotted line.
Position along the slit is shown by distance from the nucleus in arcsec, and right (positive) side is toward NW.
In all panels, positions of the nucleus and local H$\alpha$ flux minima at $R_\mathrm{minor}=-2\farcs 8$ and $+3\farcs 6$ are marked with vertical dotted lines for positional reference.
\label{minor_axis}}
\end{figure*}

\begin{figure*}
\plotone{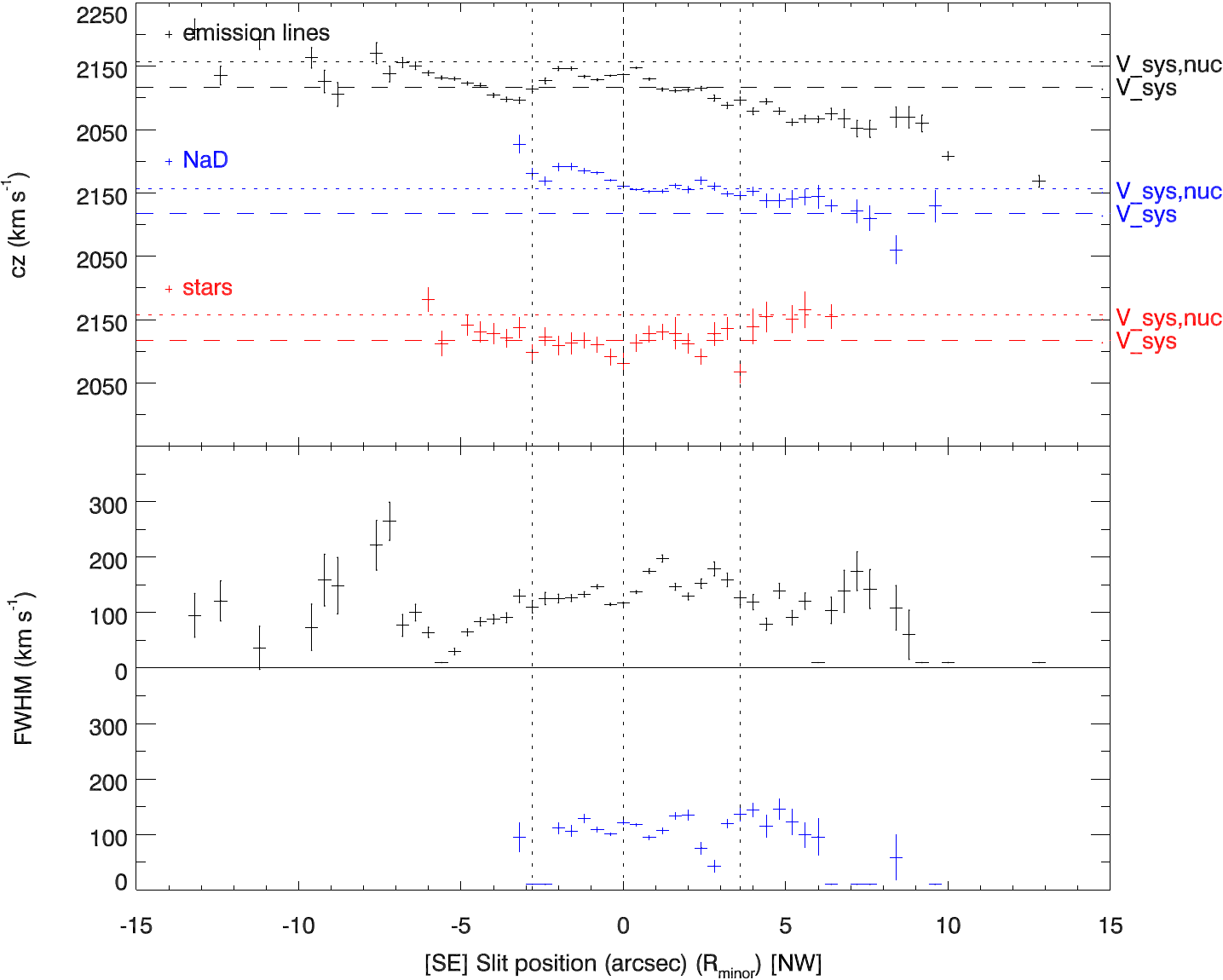}
\caption{Spatial distributions of kinematical properties along the disk minor axis.
(Top) recession velocities of the emission lines (black), Na~D absorption, (blue), and stars (red) with offsets for clarity of the figure.
The systemic velocity and the velocity origin of the outflow ($V_\mathrm{gas;~nuc.}$; see text) are marked with horizontal broken and dotted lines, respectively, for individual velocity components with corresponding colors and offsets.
(Bottom) line widths (FWHM corrected for the instrumental resolution) of the emission lines (black) and Na~D absorption (blue).
When the emission lines are unresolved, their widths are shown slightly above 0 km~s$^{-1}$ to avoid overlapping on the panel boxes for clarity of the figure.
The same positional reference lines (vertical dotted lines) as in Figure~\ref{minor_axis} are also shown.
\label{minor_axis_k}}
\end{figure*}

\begin{figure*}
\epsscale{0.9}
\plotone{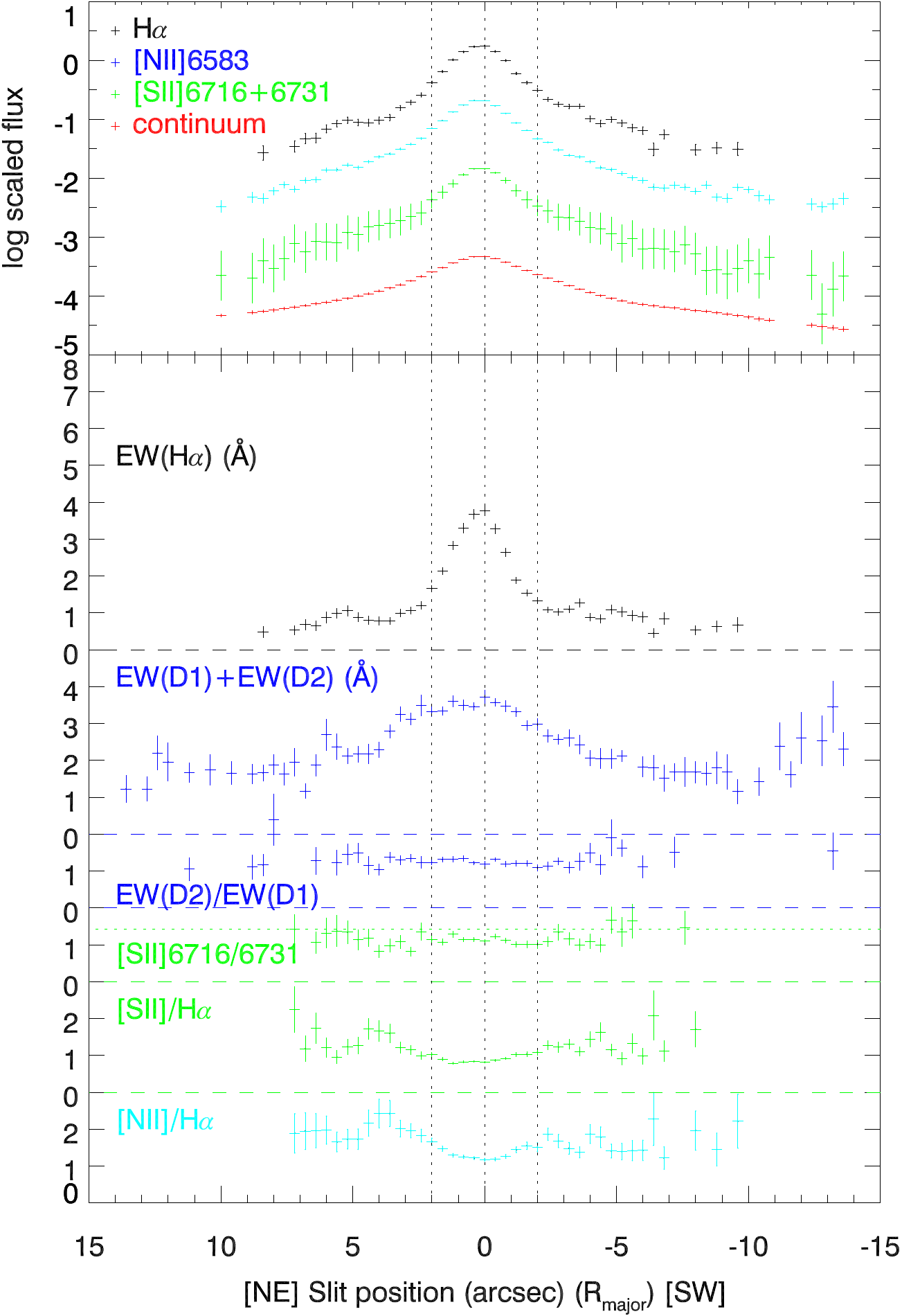}
\caption{The same plot as Figure~\ref{minor_axis}, but along the disk major axis.
Left (positive) side is toward NE.
Positions of the nucleus and outer boundaries of the circumnuclear region with enhanced H$\alpha$ EW are marked with vertical dotted lines for positional reference.
\label{major_axis}}
\end{figure*}

\begin{figure*}
\plotone{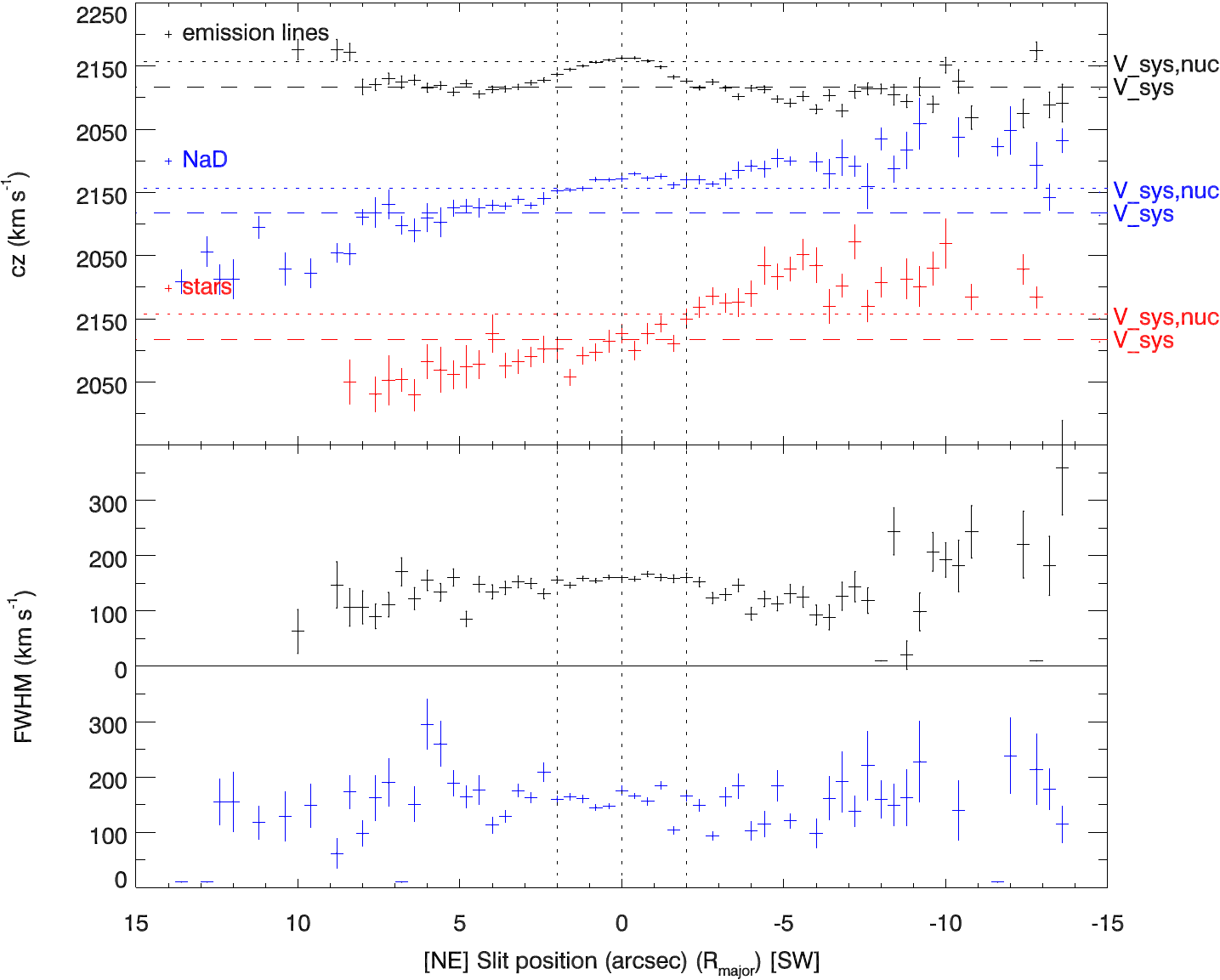}
\caption{The same plot as Figure~\ref{minor_axis_k}, but along the disk major axis.
Right (positive) side is toward NE.
The same positional reference lines (vertical dotted lines) as in Figure~\ref{major_axis} are also shown.
\label{major_axis_k}}
\end{figure*}

\begin{figure*}
\epsscale{0.9}
\plotone{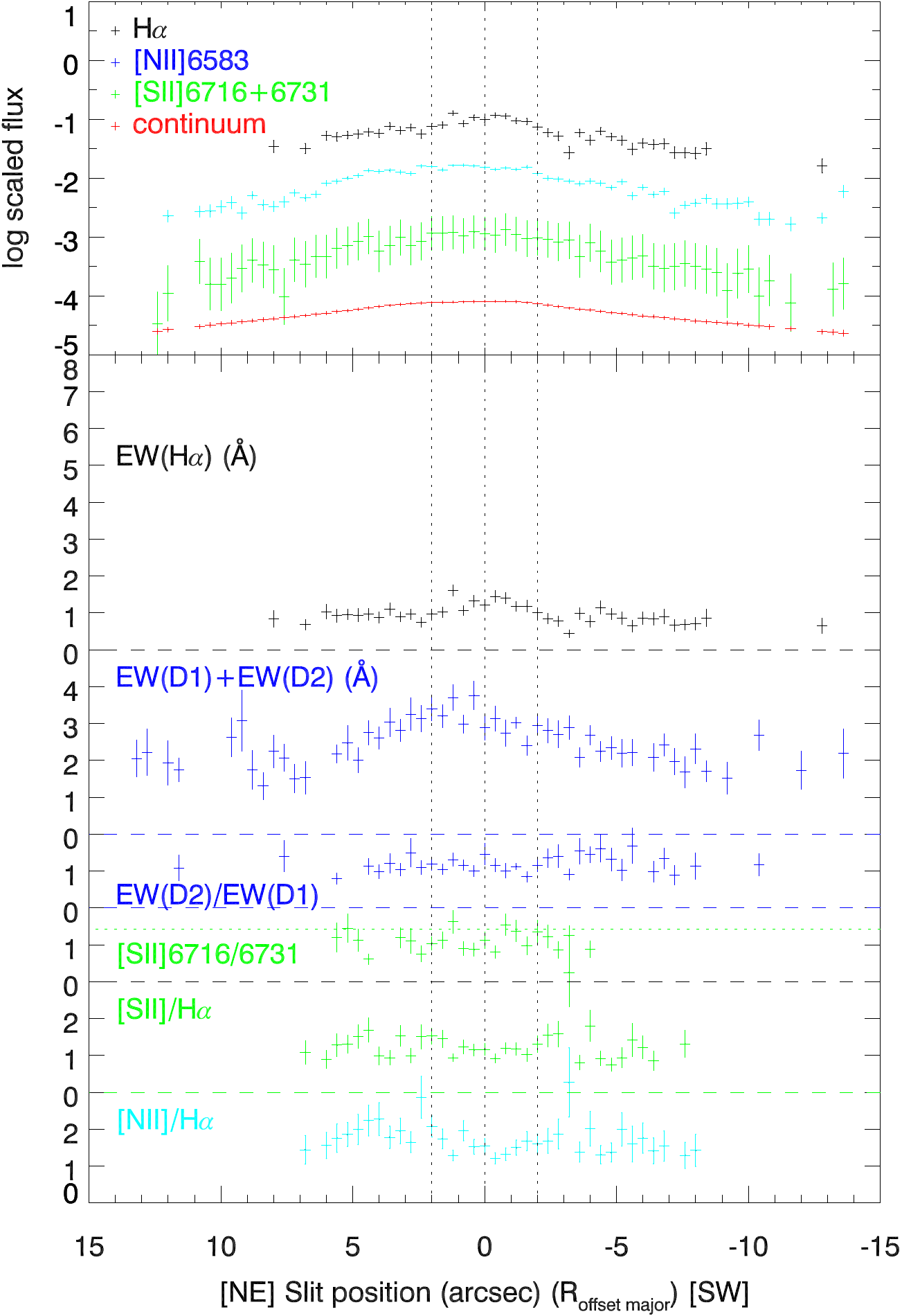}
\caption{The same plot as Figure~\ref{minor_axis}, but along the offset-major axis slit.
Left (positive) side is toward NE.
The same positional reference lines (vertical dotted lines) defined along the major axis slit (Figure~\ref{major_axis}) are also shown.
\label{offset_major_axis}}
\end{figure*}

\begin{figure*}
\plotone{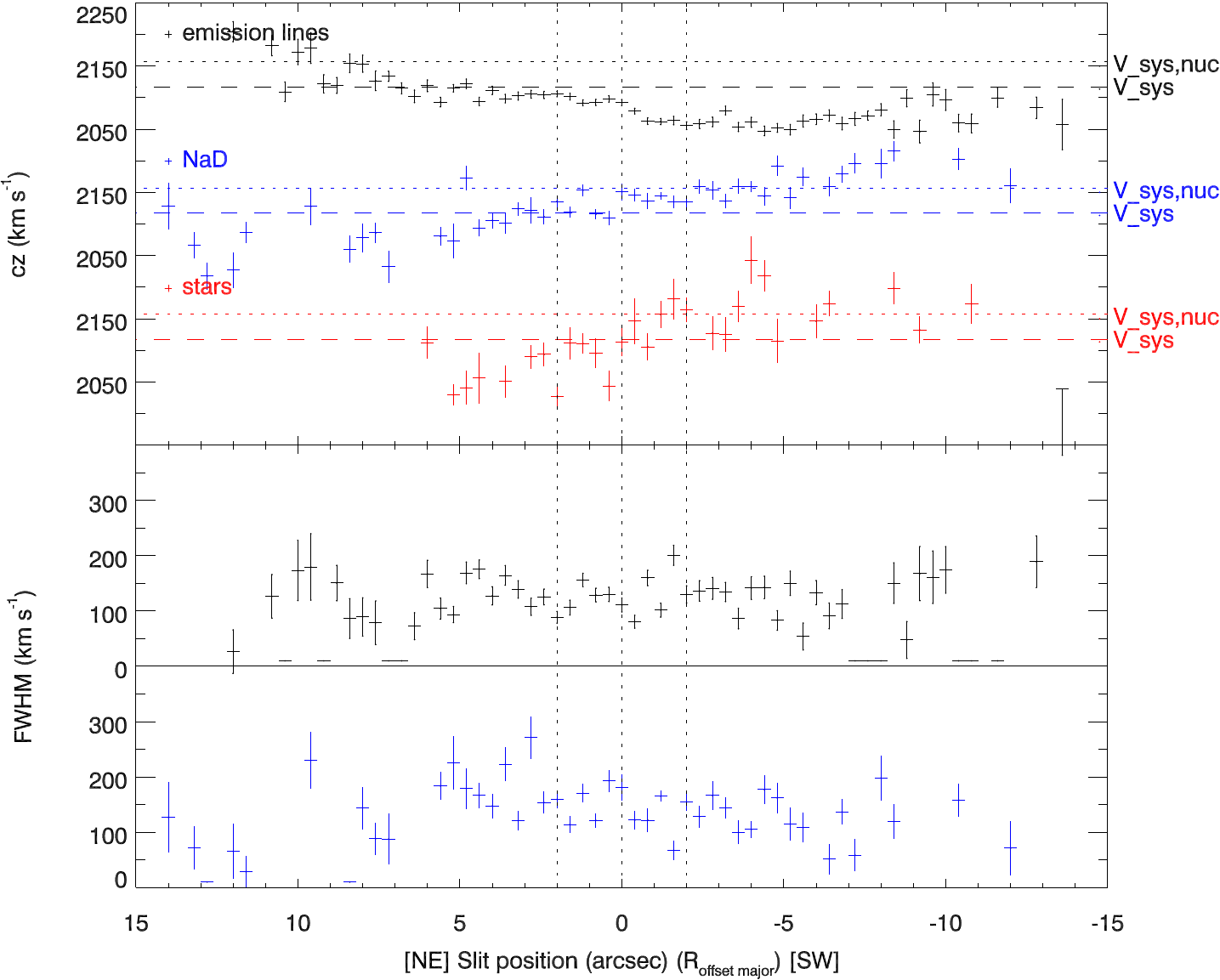}
\caption{The same plot as Figure~\ref{minor_axis_k}, but along the offset-major axis slit.
Right (positive) side is toward NE.
The same positional reference lines (vertical dotted lines) defined along the major axis slit (Figure~\ref{major_axis}) are also shown.
\label{offset_major_axis_k}}
\end{figure*}

We extracted spatial distribution of the following quantities of the ionized-gas, cold-gas, and stellar components along each of the three slits to analyze their physical and kinematical properties.
For the ionized-gas component, we measured H$\alpha$ flux, equivalent width (EW) of H$\alpha$, flux ratios of [N~{\sc ii}]$\lambda 6583$/H$\alpha$ (hereafter, [N~{\sc ii}]/H$\alpha$), [S~{\sc ii}]$\lambda 6716$+$\lambda 6731$/H$\alpha$ (hereafter, [S~{\sc ii}]/H$\alpha$), and the [S~{\sc ii}] doublet ([S~{\sc ii}]$\lambda 6716$/$\lambda 6731$), as well as their common redshift and line width.
For the cold-gas component, we measured EWs of Na~D1 and D2,
as well as their common redshift and line width.
For the stellar component, we measured EWs of stellar H$\alpha$ and Ca~{\sc i}+Fe~{\sc i} absorptions and line width of Ca~{\sc i}+Fe~{\sc i} absorption, as well as their common redshift.
The instrumental line width was subtracted from all measured widths.
At each spatial position, we measured S/N of important quantities for our analysis (fluxes of H$\alpha$ or [N~{\sc ii}] (stronger one), line width, and redshift for the ionized gas; total absorption fluxes of Na~D1 and D2, and the line width for the cold gas; continuum flux and redshift for the stars), and adopted the measurements of that position when those S/N are larger than 3.
Figures~\ref{minor_axis} to \ref{offset_major_axis_k} show most important quantities to characterize physical and kinematical properties of the three components.

\subsubsection{Systemic velocity}\label{systemic_velocity}

Stellar redshift at the nucleus is not consistent with those of the ionized- and cold-gas components.
We found that both ionized and cold gases are redshifted to $cz\simeq 2150$ km~s$^{-1}$ at the nucleus in both minor and major axis slits (Figures~\ref{minor_axis_k} and \ref{major_axis_k}).
The stellar velocity field along the disk major axis around the nucleus indicates simple solid rotation pattern of an inner galactic disk, with $cz\simeq 2100$ km~s$^{-1}$ at the nucleus (Figure~\ref{major_axis_k}; Section~\ref{results_stellar_abs}).
The SDSS nuclear spectrum also shows the same discrepancy in the redshifts.
\cite{ga12} noted, citing private communication of D. Rupke, the redshifts of $cz_\mathrm{emi}=2160$ km~s$^{-1}$ and $cz_\mathrm{star}=2100$ km~s$^{-1}$ for the emission lines and stellar spectrum, respectively.
We confirmed this by using our own spectral fitting on the SDSS spectrum.
We used the same fitting program and the fitting conditions (e.g., wavelength coverage) adopted for our FOCAS spectra to fit the SDSS spectrum, and found $cz_\mathrm{emi;~SDSS}=2154\pm 2$ km~s$^{-1}$ and $cz_\mathrm{Na~D;~SDSS}=2167\pm 3$ km~s$^{-1}$ for the emission lines and Na~D absorption, respectively.
We then measured the stellar redshift by fitting the Calcium triplet at 8500--8800\AA.
This triplet feature is modeled with triple Gaussian functions having the same redshift and line width, but with different absorption amplitudes.
We found a redshift of $cz_\mathrm{star;~SDSS}=2117\pm 4$ km~s$^{-1}$.
We note that individual redshift measurements might suffer from velocity offset caused by slight displacement of the slits or fiber from the true nucleus position due to moderately strong dust extinction around the nucleus.
However, all three measurements consistently indicate that the stellar redshift is 37--50 km~s$^{-1}$ smaller than those of the ionized- and cold-gas components at the nucleus.

The nuclear stellar redshift is roughly consistent with earlier systemic velocity measurements at other wavelengths.
\cite{sakamoto13} found a systemic velocity of $V_\mathrm{sys}=2102$ km~s$^{-1}$ (converted to optical/usual definition from their quoted value of $V_\mathrm{sys}=2088$ km~s$^{-1}$ in the radio definition) on the basis of their high-resolution ($\simeq 0\farcs 5$) molecular emission-line measurement in sub-mm wavelengths around the nucleus.
\cite{costagliola13} also found $V_\mathrm{sys}=2100$ km~s$^{-1}$ with similar molecular emission lines.
\cite{varenius17} measured the systemic redshift as $2105.4\pm 10$ km~s$^{-1}$ by using the H~{\sc i} 21~cm absorption at the nucleus.
\cite{ga12} adopted $V_\mathrm{sys}=2115$ km~s$^{-1}$ by using [C~{\sc ii}]$\lambda 178\ \mu$m emission line and H$_2$O and high-lying OH absorption lines at FIR, all of which are considered to trace the nuclear component on the basis of their detailed modeling.
All these measurements are consistent with the SDSS nuclear stellar redshift, and we adopted it as the systemic velocity ($V_\mathrm{sys}\equiv cz_\mathrm{star;~SDSS}=2117$ km~s$^{-1}$) in this work.
Consequently, the redshifts of both emission lines and Na~D absorption in the optical spectra are higher than the systemic velocity by $\simeq 50$ km~s$^{-1}$.

\begin{figure*}
\plotone{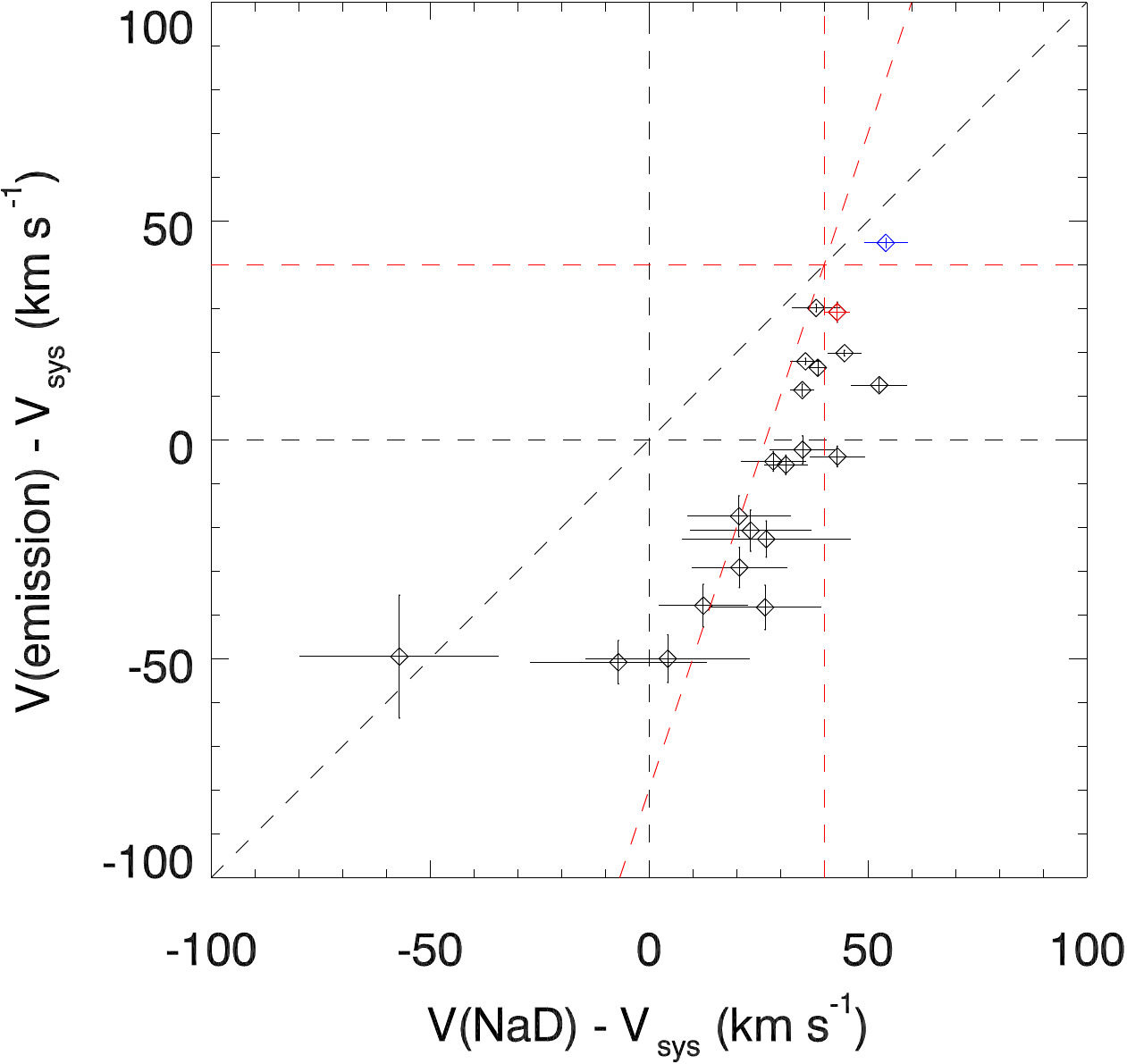}
\caption{
Velocity comparison between the Na~D absorption and emission lines along the disk semiminor axis toward NW from the nucleus.
Both velocities are shown relative to the systemic velocity ($V_\mathrm{sys}$), and error bars are for one sigma.
The velocities at the nucleus measured along the minor and major axis slits are marked in red and blue, respectively.
Black broken lines indicate the systemic velocity and the velocities when both components are in the same velocity.
Red broken lines indicate the origin of the outflow ($V_\mathrm{gas;~nuc.}=V_\mathrm{sys}+40$ km~s$^{-1}$; see text) and the velocities when the emission-line component is three times faster than the Na~D component with respect to the origin of the outflow.
\label{slit1_emi_nad_vel_comp}}
\end{figure*}

The ionized- and cold-gas components show closely correlated velocity field along the disk semiminor axis toward NW from the nucleus, and both are found at the same velocity at the nucleus about $+40$ km~s$^{-1}$ off the systemic velocity (Figure~\ref{minor_axis_k}).
Figure~\ref{slit1_emi_nad_vel_comp} directly compares their velocities to indicate the correlation with a velocity offset of $\simeq 40$ km~s$^{-1}$ from the systemic velocity at the nucleus.
The same velocity offset is also found in the measurement along the major axis slit at the nucleus.
Those findings suggest that both components belong to kinematically the same system that has a velocity offset from the systemic velocity.
As we are going to discuss in detail later (Section~\ref{dusty_superwind_model}), we attribute this system extending from the nucleus toward the disk semiminor axis toward NW to the dusty superwind outflow.
In this model, the velocity offset implies that the nuclear star cluster from which the superwind originates is kinematically distinct from the nucleus of the host galaxy.
In this work we adopted the nuclear velocity of the gaseous components as an origin of the outflow ($V_\mathrm{gas;~nuc.}\equiv V_\mathrm{sys}+40=2157$ km~s$^{-1}$).

\subsection{Spectral Synthesis of SDSS Nuclear Spectrum}\label{spectal_synthesis}

In order to study stellar population of the NGC~4418 nucleus, we performed spectral synthesis of the SDSS spectrum with STARLIGHT software (version 04; \citealt{cf05}).
This software searches for the best linear combination of single stellar population (SSP) spectra from the base library to reproduce the observed spectrum.
We adopted E-MILES SSP library (version 11.0; \citealt{vazdekis16,asad17}; see also \citealt{vazdekis10} for the original MILES SSP library) which is based on MILES stellar spectral library \citep{milesstellar}.
We combined it with the latest MILES extension for younger population ($\le 63$~Myr; \citealt{asad17}) to the previous base set of \cite{vazdekis16} ($>63$~Myr), covering 6.3~Myr--14~Gyr with 60 age bins in total.
This young extension is based on Padova94 \citep{padova94} stellar evolutionary isochrones, whereas the previous base set is based on Padova00 ones \citep{padova00}.
We adopted the SSP library built with Kroupa universal initial mass function (IMF; \citealt{kroupa01}) between 0.1--$100\ M_\mathrm{\sun}$.
We fit up to 8940\AA~to cover Calcium triplet feature at 8500--8800\AA.
This feature is strong in late-type stars and, hence, in late starburst phase when the most massive stars left the main sequence and are in red (super)giant phase, and also in much older ($\gtrsim 1$~Gyr) stellar population \citep{vazdekis03}.
Fitting this feature together with Balmer absorption, which is enhanced in post-starburst population (100~Myr--1~Gyr; e.g., \citealt{bica86b,poggianti97}), should better constrain the stellar population.
We fit the spectrum separately for different metallicities, and compared goodness of the fits for different metallicities.
We applied spectral masks for spectral features that STARLIGHT cannot fit:
emission lines of H$\beta$, [O~{\sc iii}]$\lambda 4959,5007$, [O~{\sc i}]$\lambda 6300$, H$\alpha$+[N~{\sc ii}], and [S~{\sc ii}], interstellar absorption features of Ca~{\sc ii}~H+K and Na~D, and problematic wavelengths according to the standard SDSS FITS information.

We here compare our analysis with the earlier one by \cite{shi05}.
They have already performed a very similar analysis of this galaxy by using the same SDSS data and the same (but older version of) software.
Although both analyses basically followed the standard STARLIGHT procedures, we have improved the analysis by adopting different fitting setups.
The first main difference is the adopted base SSP library.
\cite{shi05} adopted the SSP library of \cite{bc03} with STELIB stellar spectral library \citep{stelib} and Padova evolutionary isochrones, covering 5~Myr--11~Gyr with 10 age bins.
Our advantage here is much higher time resolution for the SSP evolution, and use of the MILES stellar library.
As we discuss in detail later (Sections~\ref{results_spectral_synthesis},~\ref{discussion_starburst_superwind_scenario}), higher starburst age resolution was useful to quantitatively compare the nuclear stellar population with the superwind parameters.
We also note that some general problems in STARLIGHT fitting with the STELIB stellar library have been reported earlier, whereas the fit with the MILES stellar library is known to work without these problems \citep{mateus06,asari07}.
The second main difference is the wavelength coverage.
\cite{shi05} fit the spectrum only up to 8000\AA~and did not include Calcium triplet features in the fitting for unknown reasons.
Some other minor differences are the following.
They applied spectral masks in a similar way as ours, but they included Ca~{\sc ii}~H+K in the synthesis.
They applied 3-sigma cut to actually remove some spectral pixels at Ca~{\sc ii}~K and blueward (see their Figure 1), whereas we did not apply any sigma-clipping.
They fit the spectrum with all three different metallicities at the same time during the fit .

\section{RESULTS}

\subsection{Optical Morphology and Color Distribution}\label{results_optical_image}

The $g'$-band image shows a complicated circumnuclear structure due to dust extinction (Figure~\ref{findingchart}).
Most notably, the circumnuclear structure is non-axisymmetric about the nucleus.
In particular, we found a few radial valleys in the surface brightness on the NW half of the galaxy disk at the scale of $\lesssim 3''$ from the nucleus.
The $g'-r'$ color image reveals an extended red region, forming the shape of the letter ``U'', as first reported by \cite{sakamoto13}.
The nucleus is reddest at the bottom of the ``U''.
The red region extends from the nucleus up to $\sim 4''$ toward north, and forms the two vertical sidewalls of the ``U''.
This extension is consistent with the finding by \cite{sakamoto13} based on their SDSS color map.
We confirmed that our minor axis slit goes near the middle of the ``U'' (Figure~\ref{findingchart}).
This color map shows a clear correlation with the $g'$-band image, indicating that the valleys in the $g'$-band image is due to dust extinction.
If we assume that the intrinsic color of the nucleus is the same as in the outer disk region, the peak extinction near the nucleus corresponds to $A_\mathrm{V}\simeq 1.4$ mag by using an extinction law of \cite{calzetti00}.
If the nuclear region contains intrinsically bluer stellar population as we discuss later (Section~\ref{results_spectral_synthesis}), this $A_\mathrm{V}$ estimate is a lower limit.

The $g'-r'$ color image also shows a blue, wide, fan-like region extending at the scale of $2''$--$5''$ from the nucleus between south and north east (NE) directions.
This region is about 0.1 mag (AB) bluer in $g'-r'$ than the rest of the extended disk region.
Our optical spectroscopy partially covers this blue region (Figure~\ref{findingchart}).
As we describe later (Section~\ref{results_emission_lines}), the spectra in the blue region show local excess of H$\alpha$ flux and depression of both [N~{\sc ii}]/H$\alpha$ and [S~{\sc ii}]/H$\alpha$ ratios.

\subsection{Spatial Distributions of Spectral Properties}\label{results_spatial_distribution}

\subsubsection{Stellar continuum}\label{results_cont}

Spatial distribution of the stellar continuum flux at H$\alpha$ is asymmetric about the nucleus, whereas that of the emission lines is almost symmetric, along the disk minor axis (Figure~\ref{minor_axis}).
This seems to indicate that the dust extinction works preferentially on the stellar continuum on the NW side of the disk semiminor axis.
If this is the case, the NW side of the stellar continuum is extinct at $+1''\lesssim R_\mathrm{minor}$
\footnote{$R_\mathrm{minor}$ and $R_\mathrm{major}$ are the distances from the nucleus along the minor and major axes, respectively.
Positive $R_\mathrm{minor}$ and $R_\mathrm{major}$ are on the NW and NE sides of the nucleus, respectively.
$R_\mathrm{offset~major}$ is the distance along the offset-major axis slit from the intersection with the minor axis slit.
Positive $R_\mathrm{offset~major}$ is on the NE side of the intersection.
}
$\lesssim +5''$, and the maximum continuum extinction is about 0.6 mag at $R_\mathrm{minor}\simeq +3''$.
No such asymmetry is found along the disk major axis (Figure~\ref{major_axis}).
Such dust extinction distribution is consistent with our optical color map results.

\subsubsection{Stellar absorption line}\label{results_stellar_abs}

Along the disk minor axis (Figure~\ref{minor_axis_k}), the Ca~{\sc i}+Fe~{\sc i} stellar absorption is found around the systemic velocity ($|\delta V|\lesssim 30$ km~s$^{-1}$, where $\delta V\equiv V_\mathrm{obs}-V_\mathrm{sys}$) without a clear global trend on both sides of the nucleus ($|R_\mathrm{minor}|\lesssim 6''$).
Small but systematic deviation of the stellar velocity field from the systemic velocity could be due to slight displacement of the slit from the true disk semiminor axis because of the absorption in the $R$-band nuclear image we used for the target (nucleus) acquisition on the slit during the observation.
Along the disk major axis (Figure~\ref{major_axis_k}), this absorption follows a disk-like rotation pattern around the systemic velocity, with $\delta V\simeq +130$ km~s$^{-1}$ (redshifted) on the south western (SW) side and $\delta V\sim -70$ km~s$^{-1}$ (blueshifted) on the NE side of the nucleus.
Because the velocity field can be traced further out toward SW, we adopted the projected galaxy rotation velocity as $130\pm 30$ km~s$^{-1}$, which corresponds to de-projected rotation velocity of $V_\mathrm{circ.}\simeq 150\pm 30$ km~s$^{-1}$ for $i=62^\circ$.
The dynamical mass of the disk was estimated to be $\sim 8\times 10^{9}\ M_\mathrm{\sun}$ by using this rotation velocity and a typical radius where the rotation curve is flat ($|R_\mathrm{major}|=10''$ or 1.6~kpc).
Along the offset-major axis slit (Figure~\ref{offset_major_axis_k}), the stellar velocity field is not very clear due to lower S/N, but the absorption seems to also follow the disk-like rotation similar to the one seen along the disk major axis.

Our slit spectroscopy along the major axis determined the sense of rotation of NGC~4418 for the first time to our knowledge.
The SW and NE sides are receding and approaching to us, respectively.
Given this velocity gradient, the galaxy must rotate counter-clockwise on the sky plane because the south east (SE) side of the galaxy is the near side judging from the extinction by the outflow on the NW side (Section~\ref{dusty_superwind_detail}).
Interestingly this is the opposite of the molecular gas in the central 0\farcs 5, where it is more redshifted to the NE of the nucleus and blueshifted to the SW (Section~3.4, Figures~12 and 13 in \citealt{sakamoto13}; see also Section~\ref{co_counterrotate}).

\subsubsection{Emission lines}\label{results_emission_lines}

All H$\alpha$, [N~{\sc ii}], and [S~{\sc ii}] emission lines are detected out to $\gtrsim 10''$ on both sides of the nucleus along both minor and major axes (Figures~\ref{minor_axis} and \ref{major_axis}).
They also extend along the offset-major axis slit out to $\gtrsim 10''$ from the intersection with the minor axis slit (Figure~\ref{offset_major_axis}).
Fluxes of these emission lines basically monotonically decrease with increasing distance from the nucleus, with an exception of H$\alpha$ flux along the minor axis:
H$\alpha$ flux decreases outward near the nucleus, but it then increases at $R_\mathrm{minor}\ge +3\farcs 6$ and $R_\mathrm{minor}\le -2\farcs 8$ with increasing distance from the nucleus, before it starts to decline again beyond $|R_\mathrm{minor}|\simeq 5''$.
H$\alpha$ EW is enhanced where the H$\alpha$ flux is relatively enhanced ($-7''\lesssim R_\mathrm{minor}\lesssim -3''$; $+4''\lesssim R_\mathrm{minor}\lesssim +6''$), as well as around the nuclear region ($|R_\mathrm{minor}|\lesssim 1''$).
Along the major axis slit, the EW is enhanced mostly around the nuclear region ($|R_\mathrm{major}|\lesssim 2''$), with additional slight enhancement around $R_\mathrm{major}\simeq 5''$.
Along the offset-major axis, the line flux distributions are not as clearly peaked at $R_\mathrm{offset~major}=0''$ as they are along the major and minor axes.

Both [N~{\sc ii}]/H$\alpha$ and [S~{\sc ii}]/H$\alpha$ flux ratios are enhanced ($\sim 1$--2) when compared to typical star-forming regions (e.g., \citealt{allen08}) at most regions where the emission lines are detected, with the following characteristic variations.
Along the minor axis, they are enhanced ($\sim 1$--2) within central $|R_\mathrm{minor}|\lesssim 4''$, and show local maxima ($\gtrsim 2$) at $R_\mathrm{minor}\simeq -3''$ and $R_\mathrm{minor}\simeq +4''$ near the minima of the H$\alpha$ flux and the EW.
The ratios then become relatively lower ($\simeq 0.5$--1.0) outside of this central region as the H$\alpha$ flux enhances.
Figure~\ref{focas_profmap}c shows the spectrum at a location where the line ratios are lower on the SE side of the nucleus ($R_\mathrm{minor}=-6''\pm 0\farcs 4$).
Along the major axis, both ratios are locally lower ([N~{\sc ii}]/H$\alpha\simeq 1.5$; [S~{\sc ii}]/H$\alpha\simeq 1.0$) at $R_\mathrm{major}\simeq 5''$ where both the H$\alpha$ flux and EW are locally enhanced.
We note that the regions with lower line ratios along both the minor and major axes roughly correspond to the extended blue region found in our color map (Section~\ref{results_optical_image}; Figure~\ref{findingchart}).
The [S~{\sc ii}] doublet ratio is 1.0--1.4 almost everywhere along the slits, and is $\simeq 1.1$ around the nucleus ($|R_\mathrm{minor}|\lesssim 1''$; $|R_\mathrm{major}|\lesssim 1''$).

The emission-line velocity field is complex when compared to the stellar velocity one.
Around the nucleus the emission-line velocity is $\simeq +40$ km~s$^{-1}$ off the systemic velocity in both minor and major axis slits (Figures~\ref{minor_axis_k} and \ref{major_axis_k}; see Section~\ref{systemic_velocity} for more discussion).
Along the minor axis, the emission lines monotonically blueshift with respect to the emission-line velocity at the nucleus with increasing distance from the nucleus toward NW, reaching $\Delta V\simeq -90$ km~s$^{-1}$ (where $\Delta V\equiv V_\mathrm{obs}-V_\mathrm{gas;~nuc.}$) near the NW tip of the nebula ($R_\mathrm{minor}\simeq +10''$).
Figure~\ref{focas_profmap}b shows the spectrum near the tip of the emission-line nebula ($R_\mathrm{minor}=+8''\pm 1\farcs 0$), clearly showing that the blueshifted emission lines there are [N~{\sc ii}]-enhanced compared to the lines at the nucleus ($R_\mathrm{minor}=0''\pm 0\farcs 4$; Figure~\ref{focas_profmap}a).
Along the major axis, the velocity outside of the nuclear region ($|R_\mathrm{major}|\gtrsim 2''$) is near the systemic one, i.e., it is blueshifted by $\simeq 40$ km~s$^{-1}$ with respect to the nucleus.
Such velocity field shows a clear contrast to the stellar velocity field, which shows a simple disk rotation pattern along the major axis (Section~\ref{results_stellar_abs}).
Figure~\ref{focas_profmap}d shows the spectrum in the outer disk region on the SW side of the nucleus ($R_\mathrm{major}=-5''\pm 0\farcs 4$), showing the [N~{\sc ii}]-enhanced blueshifted emission lines together with the redshifted stellar absorption.
Along the offset-major axis, the emission lines are mostly blueshifted with respect to the emission-line velocity at the nucleus, and is consistent with the measurement along the minor axis slit at $R_\mathrm{minor}=+4''$ ($\Delta V\simeq -70$ km~s$^{-1}$; Figure~\ref{minor_axis_k}), where the two slits overlap each other.
It is more blueshifted toward SW up to $\Delta V\simeq -100$ km~s$^{-1}$ (Figure~\ref{offset_major_axis_k}).
The emission lines are mostly resolved (FWHM$\simeq 100$--200 km~s$^{-1}$) everywhere along all slits, excluding at $R_\mathrm{minor}\simeq -5''$ (almost unresolved) where both [N~{\sc ii}]/H$\alpha$ and [S~{\sc ii}]/H$\alpha$ ratios show local minima ($\sim 0.5$).

\subsubsection{Na~D absorption}\label{results_nad}

The nucleus shows the deepest Na~D absorption ($EW\simeq 4.5$\AA~at $R_\mathrm{minor}\simeq 0''$; $\simeq 4$\AA~at $|R_\mathrm{major}|\lesssim 1''$; Figures~\ref{minor_axis} and \ref{major_axis}).
The enhanced absorption extends asymmetrically toward NW along the disk minor axis ($EW=3$--4\AA~out to $R_\mathrm{minor}\simeq +5''$).
Toward the opposite direction (SE), the EW rapidly decreases outward, and no absorption is detected beyound $R_\mathrm{minor}=-4''$.
We note that, beyond the region with the Na~D absorption, we found a possible blueshifted Na~D emission at $R_\mathrm{minor}=-6''\pm 0\farcs 4$ (Figure~\ref{focas_profmap}c).
Along the disk major axis, the absorption distributes more symmetrically around the nucleus and becomes shallower outward on both sides of the nucleus ($EW\gtrsim 3$\AA~at $R_\mathrm{major}=-2''$ -- $+3''$; $EW\simeq 2$\AA~at $|R_\mathrm{major}|\gtrsim 5''$).
Along the offset-major axis slit, the EW is $\sim 3$\AA~near the intersection with the minor axis slit, and it decreases outward (Figure~\ref{offset_major_axis}).
The Na~D doublet ratio is 1.0--1.2 everywhere along all slits.

The velocity field of the Na~D absorption is also complex when compared to both the emission lines and the stellar velocity fields.
The Na~D absorption at the nucleus is found at the emission-line velocity at the nucleus, being offset from the systemic velocity (Figure~\ref{minor_axis_k}; Section~\ref{systemic_velocity}).
Along the minor axis, it monotonically blueshifts with increasing distance from the nucleus toward NW, as the emission lines do, reaching $\Delta V\simeq -50$ km~s$^{-1}$ near the NW tip of the nebula ($R_\mathrm{minor}\simeq +8''$).
Figure~\ref{focas_profmap}b shows such blueshifted absorption at $R_\mathrm{minor}=+8''\pm 1''$ when compared to the nucleus (Figure~\ref{focas_profmap}a).
We note that the blueshift of the absorption is only $\simeq 1/3$ of that of the emission lines at the same position (Figure~\ref{slit1_emi_nad_vel_comp}).
Along the major axis, the absorption shows a systematic global velocity gradient, redshifting and blueshifting on the SW and NE sides of the nucleus, respectively, up to $|\Delta V|\sim 100$ km~s$^{-1}$ (Figure~\ref{major_axis_k}).
This trend is similar to the disk rotation seen in the stellar velocity field in the same slit (excluding the vicinity of the nucleus where it shows offset; see above), while it is quite different from the velocity field of the emission lines (Section~\ref{results_emission_lines}).
Along the offset-major slit, the absorption shows a small but systematic global velocity gradient up to $|\Delta V|\sim 50$ km~s$^{-1}$ in the same sense of the stellar velocity field (Figure~\ref{offset_major_axis_k}) while showing a clear contrast to the emission-line velocity field (Section~\ref{results_emission_lines}).
The absorption line is mostly resolved (FWHM$=100$--150 km~s$^{-1}$) everywhere along all slits.

\subsubsection{Counter-rotation of the molecular gas core and the plasma outflow}\label{co_counterrotate}

The velocity gradients of the molecular gas in the central 0\farcs 5 and the galaxy disk are approximately parallel but have opposite directions (Section~\ref{results_stellar_abs}).
The molecular gas must be in counterrotation with respect to the main stellar disk of the galaxy if the gas rotates in the galactic plane.
Or the nuclear gas disk must be tilted by more than about $60\degr$ from the galactic plane in order for it to show the reversed velocity gradient while having its angular momentum vector and that of the galaxy disk pointed in the same hemisphere.
Another possible interpretation of the misaligned velocity gradient of the molecular gas is that the gas has significant non-rotational velocities.
In either case, the anomalous velocity structure of molecular gas in the nuclear region indicates ongoing or recent disturbance of signifiant magnitude there.

The superwind that we will discuss further in Section~\ref{dusty_superwind_model} appears to rotate in the same direction as the central molecular gas concentration does.
Along the offset-major axis (Figure~\ref{offset_major_axis_k}), emission lines from ionized gas show a velocity gradient that is more blueshifted in the SW and redshifted in the NE;
the direction of the velocity gradient is the same as that of the molecular gas near the center.
In contrast, stars and Na~D lines along the offset-major axis show the opposite sense of velocity gradient with respect to the ionized gas and molecular gas, and the same sense of velocity gradient as the galactic disk does along the major axis.
This velocity gradient is as expected for the stellar lines because the stellar light along the offset-major axis must be from the galactic disk behind the superwind; we expect little stars in the superwind.
The reason why the Na~D absorption shows the same pattern of rotation as the disk stars does is probably that the disk ISM behind the superwind contributes significantly to the observed Na~D feature.
We regard the ionized gas a better tracer of the kinematics of the superwind plasma than the Na~D line at least on this offset-major axis.

\subsection{Nuclear Stellar Population}\label{results_spectral_synthesis}

\begin{figure*}
\plotone{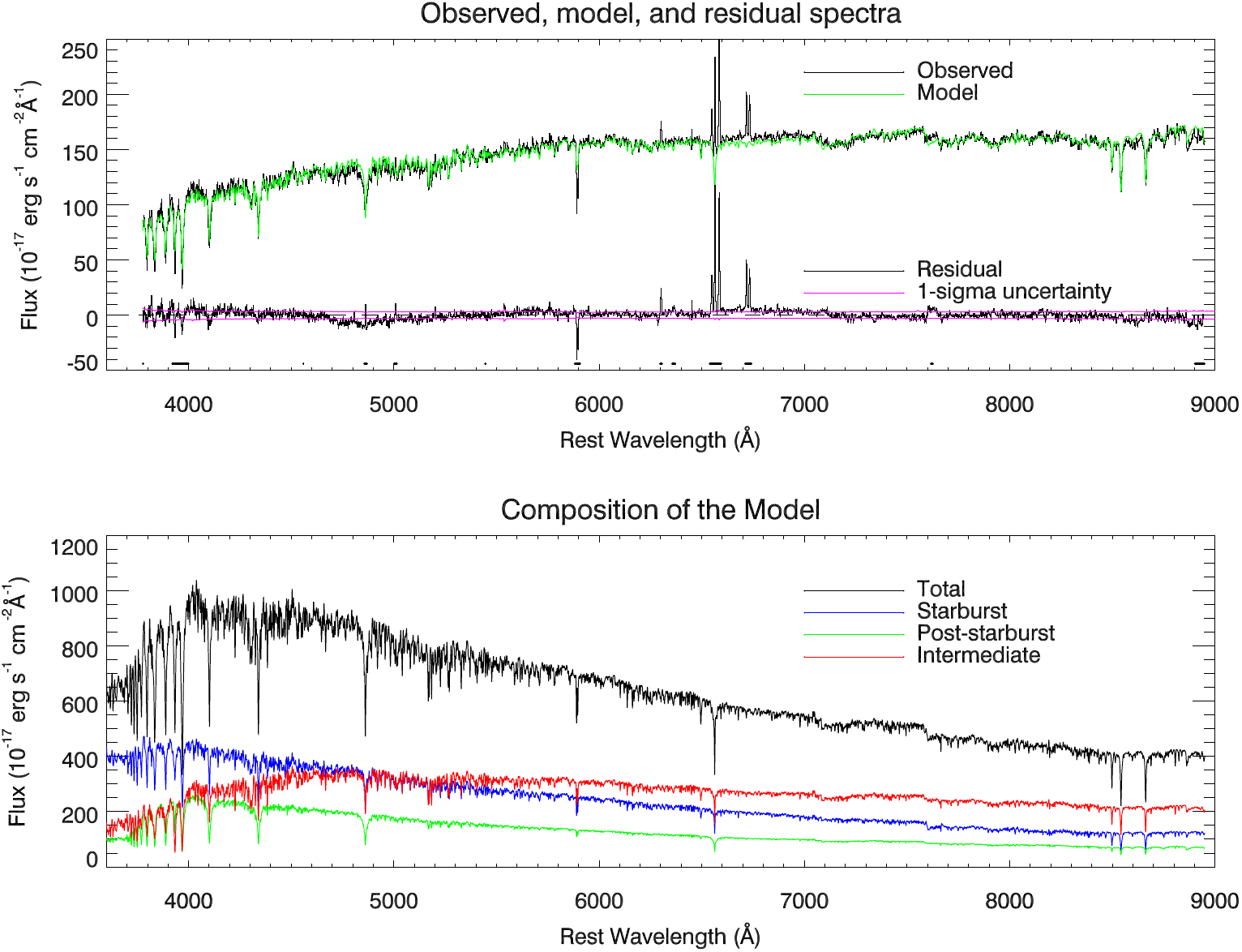}
\caption{The SDSS spectrum of NGC~4418 nucleus and the result of the spectral synthesis.
(Top panel) the STARLIGHT best-fit model spectrum (green) is overlaid on the observed spectrum corrected for Galactic reddening (upper black).
Both intrinsic extinction and velocity broadening due to both instrumental resolution and the stellar velocity dispersion are applied to this model spectrum.
The residual (observed$-$model) spectrum (lower black) is also shown around zero flux level.
Short black horizontal bars near the bottom of this panel indicate masked wavelength regions during the fit.
Two purple lines overlaid on the residual spectrum show $\pm 1$-sigma uncertainty of the observed spectrum to indicate significance of the residual spectrum.
(Bottom panel) the SSP populations that constitute the best-fit model spectrum.
The intermediate (1.4~Gyr), post-starburst (300--600~Myr), burst (8~Myr), and total of all these populations are shown in red, green, blue, and black, respectively.
Both Galactic and intrinsic extinctions and the velocity broadening are not applied to these spectra.
\label{sdss_starlight_spec}}
\end{figure*}

Figure~\ref{sdss_starlight_spec} top panel shows the spectral synthesis result of the NGC~4418 nucleus.
The best-fit spectrum (reduced $\chi^2=1.35$) was found with the solar metallicity ($[Z/Z_\mathrm{\sun}]=+0.06$) SSP library, an intrinsic extinction of $A_\mathrm{V}=1.76$ mag, and total (sum of all SSPs) initial stellar mass of $9.9\times 10^8\ M_\mathrm{\sun}$ within $3''$ diameter aperture.
This extinction is consistent with our earlier measurement based on the $g'-r'$ color map ($A_\mathrm{V}\gtrsim 1.4$ mag; Section~\ref{results_optical_image}).
For comparison, we also fit with the same wavelength coverage as \cite{shi05} (only up to 8000\AA, and without masking Ca~{\sc ii}~H+K), and found almost the same fit (reduced $\chi^2=1.37$).
This confirms that our fit shows significant improvement from the result of \cite{shi05} (reduced $\chi^2=2.47$) thanks to the improved SSP library we adopted (Section~\ref{spectal_synthesis}).
We note that our best reduced $\chi^2$ is still larger than 1, indicating that the fit is still not statistically satisfactory.
In fact, the residual (observed$-$model) spectrum (excluding masked features) shows some systematic pattern changing smoothly at the typical scale of several 100\AA, and it sometimes becomes larger or smaller than $\pm 1$ sigma uncertainty of the input SDSS spectrum.
Besides such large scale residual features, the fit reproduces most absorption features that are typically seen in extragalactic spectra, such as Balmer (excluding H$\alpha$ and H$\beta$, which were masked in our fit), G band, Mg~Ib, and Calcium triplet features, very well.
This residual spectrum pattern persists when we instead adopt SSPs with different metallicities or IMF (Chabrier IMF; \citealt{chabrier03}), or libraries including a power-law continuum ($f_\nu\propto\nu^{-1.5}$) component.
The fit with the power-law component was to test possible AGN contribution, and we found it negligibly small.
Therefore, we confirmed that the fit with solar metallicity SSPs with the Kroupa universal IMF is practically the best, and adopted this fit in the following analysis and discussions.

The observed nuclear spectrum can be reproduced by only a small number of SSP components: starburst ($8\pm 1$~Myr), post-starburst (300--600~Myr), and intermediate populations ($1.4\pm 0.2$~Gyr).
Here, because we found only one SSP component for the starburst and intermediate populations, we found their ranges of the burst ages by considering ages of the adjacent SSP components.
Their relative flux contributions at 4050\AA~are 41\%, 23\%, and 36\% for the respective components.
Their initial stellar masses are $\simeq 10^7\ M_\mathrm{\sun}$, $\simeq 10^8\ M_\mathrm{\sun}$, and $\simeq 10^9\ M_\mathrm{\sun}$ for the respective populations within the fiber aperture of 500~pc in diameter.
Therefore, our spectral synthesis result strongly indicates intermittent star formation activity in the nucleus of NGC~4418.
We note that \cite{cf05} (and references therein) recommend to bin over spectrally similar components for robust description of the spectral synthesis results, because noise in the observed spectrum washes away the differences between such components.
Even so, our result seems to strongly indicate two short-duration starbursts (the starburst and post-starburst components) on top of the intermediate population at the nucleus, because the activity between the small number of SSP components is found to be negligibly low.
We caution, however, that the possible ranges of the burst age directly derived from the synthesis (e.g., $8\pm 1$~Myr for the starburst population) are subject to intrinsic limitations of the method and our setup (e.g., available spectral features to be synthesized, the adopted base library for the synthesis, and noise in the observed spectrum).
The starburst and intermediate populations dominate the observed spectrum at bluer and redder wavelengths than $\simeq 5000$\AA, respectively (Figure~\ref{sdss_starlight_spec} bottom panel).
The intermediate population dominates Ca~{\sc ii}~H+K, G band, Mg~Ib, and complex absorption features at $>7000$\AA~including the Calcium triplet feature.
The post-starburst population contributes most to the Balmer absorption (e.g., \citealt{bica86b,poggianti97}).
The very old component ($>7$~Gyr), which is usually a major contributor to the total flux and mass in the bulge of Sa-type galaxies, contributes negligibly small.
We also note that the extinction inferred from this spectral synthesis was derived by assuming a screen geometry of the dust layer, giving lower extinction when compared to the case of mixed distribution of dusts and stars.
Therefore, this extinction is a lower limit.

We need robust constraint on ages and masses of the starburst component to examine the starburst scenario of this nucleus (Section~\ref{discussion_starburst_superwind_scenario}).
In particular, we want to test if this starburst is compatible with a compact star cluster postulated within the sub-mm core by \cite{sakamoto13} for the case of no luminous AGN (hereafter, the core cluster; Section~\ref{discussion_where_is_starburst}).
Unfortunately, the youngest SSP component in our synthesis base is 6.3~Myr old (Section~\ref{spectal_synthesis}), and we cannot test if the core cluster with age of a few~Myr ($\lesssim 5$~Myr) can fit the observation.
The spectral shape of a starburst does not change much in time around the age of several$-$10~Myr at 3500--4500\AA, where the starburst dominates over other older components in our spectral synthesis.
Because OB stars that dominate such blue wavelength region in the starburst do not show prominent characteristic absorption features there, we cannot easily constrain the starburst age if we allow variable optical reddening among SSP components to absorb small changes in the spectral slope.
Therefore, we here simply calculate range of the stellar masses of the starburst component given a possible range of the burst ages around the best-fit age of 8~Myr.
The mass is scaled from the best-fit case ($1\times 10^7\ M_\mathrm{\sun}$; see above) based on different mass-to-luminosity ratios for different burst ages, according to the same {\it starburst99} model used in Section~\ref{consistency_starburst_superwind}.
We found the stellar masses to be $0.5\times 10^7\ M_\mathrm{\sun}$, $1.0\times 10^7\ M_\mathrm{\sun}$, and $1.4\times 10^7\ M_\mathrm{\sun}$ for the starburst of 5~Myr, 8~Myr, and 10~Myr, respectively.
Here, the age of 5~Myr is chosen as a lower limit, because such a young starburst over-predicts the H$\alpha$ luminosity (Section~\ref{scenario_consistency_ha_ir}).
The upper-limit age of 10~Myr is based on the dynamical time argument of the superwind development (Section~\ref{dusty_superwind_params}).

We measured reddening on the ionized gas at the nucleus by using the Balmer decrement method.
We measured both H$\alpha$ and H$\beta$ emission line fluxes by utilizing the synthesized stellar spectrum for subtracting the underlying stellar spectrum.
We found the H$\alpha$/H$\beta$ emission-line flux ratio $\simeq 8.7$, which is similar to the result of \cite{shi05} (6.42).
This ratio is significantly higher than the unabsorbed ratio of 2.8 for H~{\sc ii} regions \citep{agn2} or 2.9--3.2 for shock-excited nebulae \citep{allen08}, and extinction of $A_\mathrm{V}=3.5$--3.9 mag is required by assuming the extinction law of \cite{calzetti00}.
This extinction is larger than that on the stellar continuum as inferred from the spectral synthesis ($A_\mathrm{V}=1.76$ mag or larger).
This difference can be explained if the ionized gas is from a nuclear region that is more compact that the SDSS fiber size ($3''$ in diameter), and the extinction is larger closer to the nucleus.
We indeed found such extinction distribution in our optical color map (Section~\ref{results_optical_image}).

The multiple starburst episodes we found is qualitatively consistent with the result of \cite{shi05}.
To enable direct comparison with their result, we rebinned our results into their three population groups: young ($<25$~Myr), intermediate (25~Myr--1~Gyr\footnote{\cite{shi05} defined their intermediate population age range as 25~Myr--100~Myr according to their text.
However, we believe that the correct range is 25~Myr--1000~Myr.
This is because the tabulated flux contribution numbers in their Table 1 match with their histogram representation in their Figure 1 only with this revised range.}), and old ($>1$~Gyr) population groups.
Here, we tweaked their old population age range from $>2.5$~Gyr to $>1$~Gyr, because we have several SSP components within 1--2.5~Gyr, while \cite{shi05} have no component there.
We found 41\%, 23\%, and 35\% flux contributions at 4540\AA~in their young, intermediate, and old population groups, respectively, with an extinction of $A_\mathrm{V}=1.76$ mag.
\cite{shi05} found 19.5\%, 24.5\%, and 56.0\% flux contributions at the same wavelength in these respective population groups with an extinction of $A_\mathrm{V}=1.32$ mag.
When compared to \cite{shi05}, our model consists of younger and intrinsically bluer population that suffers more reddening.
Both models are consistent in the sense that this galaxy nucleus has experienced multiple starbursts.
\cite{shi05} found two age ranges of little star formation at $\sim 100$~Myr and $\sim 1$~Gyr.
We found similar star formation gaps at 10--300~Myr and 700~Myr--1.2~Gyr.

The star-formation activities in NGC~4418 might have been triggered by an interaction with its neighbor galaxy VV~655 located at $\simeq 32$~kpc ($3\farcm 2$) away on the sky \citep{evans03,varenius17}.
\cite{varenius17} discovered an H~{\sc i} 21~cm bridge connecting the two and argued that they first passed each other $\sim 190$~Myr ago.
The post-starburst population we found ($\simeq 300$--600~Myr old; Section~\ref{results_spectral_synthesis}) could have been triggered by this interaction.

\section{PHYSICAL CONDITIONS OF THE EXTENDED GASEOUS COMPONENTS}\label{basic_characteristics}

\subsection{Ionized Gas}\label{analysis_nebula}

The ionized-gas nebula has a projected size of $\simeq 10''$ (1.7~kpc) along the disk semiminor axis.
\cite{lehnert95} reported a half light radius of 1.0~kpc (after converted to our assumed distance) on the basis of their narrow-band H$\alpha$+[N~{\sc ii}] image, being roughly consistent with our spectroscopic measurement.
The [S~{\sc ii}] doublet ratio is $\simeq 1.0$ around the nucleus and 1.0--1.4 along the minor axis slit.
The doublet ratios of 1.0 and 1.43 correspond to the local electron density of $\simeq 500$ cm$^{-3}$ and a limit for the low-density condition, respectively \citep{agn2}.
An RMS electron density averaged along the minor axis slit was estimated to be $\sim 1$ cm$^{-3}$ by using the H$\alpha$ luminosity and the line-of-sight length of the nebula.
We assumed the line-of-sight length to be equal to its representative width of the nebula, and adopted as the width the size of the region showing enhanced H$\alpha$ EW along the major axis slit ($4''$ or 660~pc) as the representative nebula width.
The volume filling factor of the ionized gas was then estimated to be $\sim 3.3\times 10^{-5}$ from the RMS and local electron densities, by adopting 100 cm$^{-3}$ as a representative value of the latter.

\begin{figure*}
\plotone{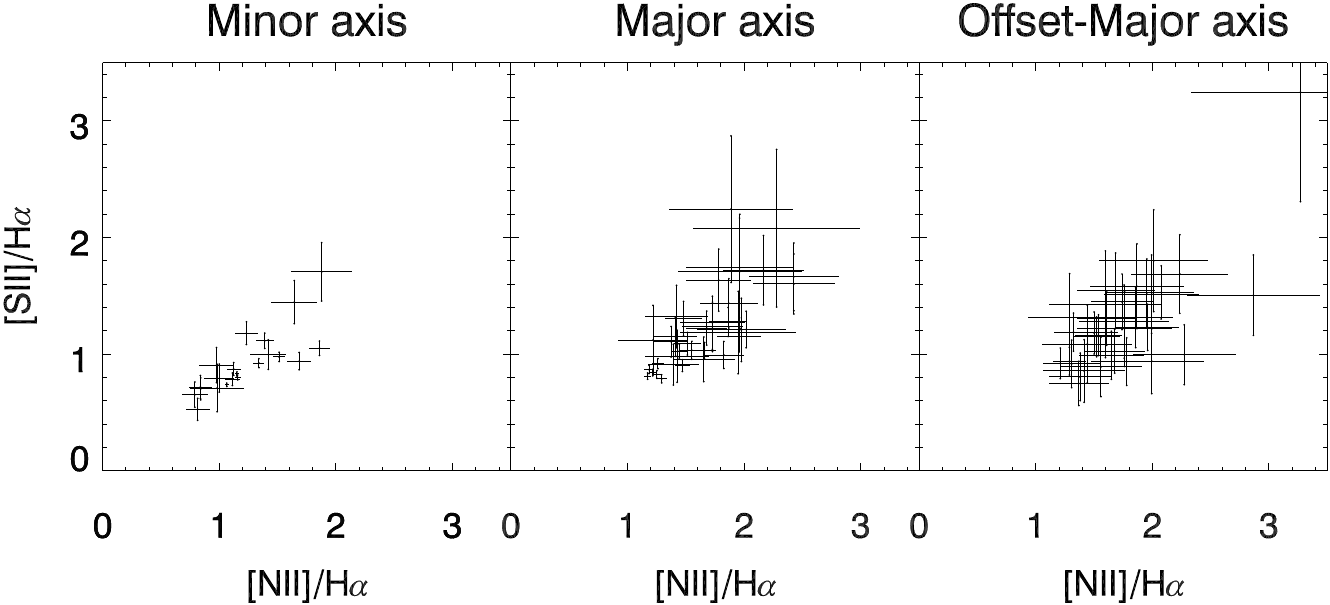}
\caption{Correlations between [N~{\sc ii}]/H$\alpha$ and [S~{\sc ii}]/H$\alpha$ line ratios.
Left, center, and right panels show data along the minor, major, and offset-major axis slits, respectively.
Error bars are for one sigma.
\label{lineratio_lineratio}}
\end{figure*}

Star formation cannot explain the observed characteristics of the ionized-gas nebula, but shock excitation can.
First, both [N~{\sc ii}]/H$\alpha$ and [S~{\sc ii}]/H$\alpha$ line ratios are typically $\gtrsim 1$ at both the nucleus and the extended region.
These ratios are much higher than the typical ones in H~{\sc ii} regions and starburst nuclei ([N~{\sc ii}]/H$\alpha<0.5$; [S~{\sc ii}]/H$\alpha<0.1$; e.g., \citealt{mccall85}).
On the other hand, the solar abundance shock model predicts [N~{\sc ii}]/H$\alpha\lesssim 1$ and $\lesssim 2$ for the shock velocities of 200 km~s$^{-1}$ and 300 km~s$^{-1}$, respectively (e.g., \citealt{allen08}), therefore, the shock can explain the observed enhanced ratios.
Second, the velocity field is quite different between the ionized gas and the stars in both minor and major axis slits, indicating that most of the off-nuclear ionized gas is not associated with the galaxy disk.
Third, clear positive correlations between [N~{\sc ii}]/H$\alpha$ and [S~{\sc ii}]/H$\alpha$ are found along the slits (Figure~\ref{lineratio_lineratio}).
Shock excitation can naturally explain such a correlation as a function of shock velocity (e.g., \citealt{allen08}).
We note, however, that the same correlation can be reproduced by having different amounts of contamination from the disk H~{\sc ii} regions at different positions along the slits.
For example, both line ratios are lower at $+4''\lesssim R_\mathrm{minor}\lesssim +6''$ and $-7''\lesssim R_\mathrm{minor}\lesssim -3''$ along the minor axis slit, and also at $R_\mathrm{major}\simeq 5''$ along the major axis slit.
Because the H$\alpha$ flux is locally enhanced in these regions, additional contribution by disk H~{\sc ii} regions likely changes the observed ratios there.
Some pieces of supporting evidence on this idea is that these regions correspond to the disk showing bluer color (Section~\ref{results_optical_image}), and show narrower emission line width.

\subsection{Cold Gas}\label{analysis_cold_gas}

An enhanced Na~D absorption around the nucleus and on the NW side of the nucleus along the disk semiminor axis is likely due to interstellar absorption by dusty cold gas.
Stellar photospheric Na~D absorption, which is enhanced in cool stars (e.g., \citealt{jacoby84}), is very unlikely to reproduce the observations for two reasons.
First, distribution of the enhanced Na~D absorption is similar to the reddened region in the optical color map (Section~\ref{results_optical_image}; \citealt{sakamoto13}).
The Na~D EW is most enhanced near the nucleus where the extinction is largest, and is also enhanced where the ``U''-shaped extinct region extends toward NW.
Second, the Na~D absorption does not follow the stellar velocity field along the disk semiminor axis.
The observed low Na~D doublet ratio ($EW_\mathrm{Na~D2}/EW_\mathrm{Na~D1}\simeq 1$) indicates that the Na~D absorption is optically thick at most of the region.
This ratio becomes $\lesssim 2$ if the absorption is optically thin, and it approaches unity as the absorption saturates when Na~{\sc i} column density is on the order of $N_\mathrm{Na~I}\sim 10^{13}$ cm$^{-2}$ or larger.
However, some fraction of the stellar continuum is left unabsorbed even at the bottom of the Na~D2 absorption feature.
This is most likely because the absorbing material does not fully cover the continuum emitting sources, i.e., the cold gas is distributed in patchy clouds and the covering factor is less than unity (e.g., \citealt{heckman00,martin05,martin06,rupke05b}).
The covering factor is typically $\sim 0.3$ in ULIRGs \citep{martin05} and 0.25--0.5 in IRGs (infrared galaxies with $L_\mathrm{IR} <10^{12}\ L_\mathrm{\sun}$; \citealt{rupke05b}).
We measured the covering factor from the minimum flux at the bottom of the D2 absorption at each spatial position along the minor axis slit, and found that it is mostly $\simeq 0.4$ at positions where $EW_\mathrm{Na~D}>3$\AA, with a peak of $\simeq 0.5$ at the nucleus.

We estimated the column density of the cold gas in the extended Na~D-enhanced region along the disk semiminor axis to be on the order of $\sim 10^{21}$ cm$^{-2}$.
By following \cite{martin06} (see \citealt{heckman00} and \citealt{rupke05a} for alternative approaches), we fit the absorption profile after correcting for the covering factor to estimate the apparent optical depth.
This apparent optical depth is based on an optically thin approximation, and is therefore a lower limit to the actual one.
We fit the continuum-normalized intensity profile of the absorption with a function of $\exp(-\tau(\lambda))$, where $\tau(\lambda)=\tau_\mathrm{D2}(\lambda)+\tau_\mathrm{D1}(\lambda)$ and $\tau_\mathrm{D2}(\lambda-\lambda_\mathrm{D2})=1.98\times\tau_\mathrm{D1}(\lambda-\lambda_\mathrm{D1})$.
Optical depth profile of each line is modeled with a Gaussian function.
We ignored contribution from stellar Na~D absorption.
We found that the optical depth is $\int\tau_\mathrm{D1}d\lambda\simeq 6$ at $-1''\lesssim R_\mathrm{minor}\lesssim +6''$, and the corresponding Na column density is $N_\mathrm{Na~I}=1.02\times 10^{13}\times\int\tau_\mathrm{D1}(\lambda)d\lambda\simeq 6\times 10^{13}$ cm$^{-2}$ \citep{martin06} there.
To estimate the total H~{\sc i} column density from the measurement of gas-phase neutral Na, we need to correct for the ionizing fraction and dust depletion factor of Na, as well as metallicity.
We found the column density of $N_\mathrm{H~I}\simeq 3\times 10^{20}$ cm$^{-2}$ and $\simeq 1\times 10^{22}$ cm$^{-2}$ by adopting ULIRG (\citealt{ferlet85}) and standard Galactic (\citealt{wakker00}; see also \citealt{martin05,martin06}) conditions, respectively.
Here, the large discrepancy (by a factor of 30) between the two cases is due to difference in the ionization fraction correction.
Na is very easily ionized because of its very small ionization potential (5.1~eV), and the neutral fraction is expected to be larger in ULIRGs because of better photon shielding in the dusty environment (\citealt{ferlet85}; see also \citealt{martin05,martin06}).
For comparison, \cite{sakamoto13} reported a mean column density of $N_\mathrm{H+H2}=1.3\times 10^{21}$ cm$^{-2}$ in the reddest 1~kpc$^2$ region by using of the optical extinction, being broadly consistent with our estimate by using the Na~D absorption.
In the following, we adopt a hydrogen column density of $1.3\times 10^{21}$ cm$^{-2}$ from \cite{sakamoto13} due to the considerable uncertainties in our estimate.
Although we have no information about the molecular gas contents within this extended Na~D enhanced region, we simply use this column density as a measure of the total cold gaseous component there.

\section{THE DUSTY SUPERWIND MODEL}\label{dusty_superwind_model}

\subsection{Evidence for Dusty Superwind}\label{dusty_superwind_evidence}

Most of the characteristics of the extended nebula of both ionized and cold gases suggest a dusty superwind in NGC~4418.
The enhanced interstellar Na~D absorption shows a monotonically blueshifting velocity with increasing distance from the nucleus toward NW along the disk semiminor axis over a kpc scale (Figure~\ref{minor_axis_k}).
This blueshifted interstellar absorption on one side of the nucleus strongly indicates that the cold dusty gas is associated with a nuclear outflow located in front of the galaxy disk.
The shock-heated ionized gas also shows a monotonically blueshifting velocity with increasing distance from the nucleus toward NW along the disk semiminor axis over a kpc scale (Figure~\ref{minor_axis_k}; Section~\ref{analysis_nebula}), indicating that both cold and ionized gases are spatially and kinematically associated with each other over a galactic scale.
We note, however, that the outflow velocity of the cold gas is only $\simeq 1/3$ of that of the ionized gas (Section~\ref{systemic_velocity}).
All these pieces of information strongly suggest that both dusty cold gas and shock-heated ionized gas form a large-scale outflow pointing toward the disk polar directions, or a duty superwind.
This dusty cold gas outflow and the extended H~{\sc i} emission that \cite{varenius17} attributed to the superwind outflow are similar in the outflow direction not only on the sky (NW from the nucleus) but also in the velocity field (blueshifted from the nucleus).

For comparison, the prototypical superwind in M82 is known to show qualitatively similar characteristics as the one in NGC~4418 (e.g., \citealt{alton99,heckman00,yoshida11,contursi13,salak13,leroy15}; see Section~\ref{dusty_superwind_comp_M82} for detailed quantitative comparison).
Its outflow velocity monotonically increases outward (e.g., \citealt{bland88,mckeith95,shopbell98,greve04,greve11}) in such a way that the ionized-gas outflow is faster than those of the molecular gas, cold neutral atomic gas, and dusts (e.g., \citealt{seaquist01,yoshida11,contursi13}).
The cold-gas outflows are ubiquitously found by using Na~D absorption line in starburst galaxies including (U)LIRGs (e.g., \citealt{rupke02,martin05,martin06,rupke05b,chen10}), and their outflow velocities are generally similar or slower than the ionized-gas ones (e.g., \citealt{rupke02,rupke05b}).
Theoretical studies of the superwind (e.g., \citealt{suchkov94,strickland00}) can explain these characteristics.

We here examine a possible alternative origin of the off-nuclear material that does not belong to the galaxy disk.
\cite{varenius17} found a bridge in H~{\sc i} 21~cm emission between NGC~4418 and its companion galaxy VV~655.
They argued that various activities of the NGC~4418 nucleus is fueled through this bridge as a result of their past interaction.
However, it is not clear how this bridge is connected to the nucleus of NGC~4418 because their H~{\sc i} data have a low spatial resolution ($61\farcs 5\times 51\farcs 0$).
Although such gas supply to the nucleus might mimic the postulated superwind as described above, we found this unlikely by using the metallicity argument.
Spectra of the companion galaxy VV~665 were taken with the Robert Stobie Spectrograph on the Southern African Large Telescope.
These show emission line ratios, such as a small [N~{\sc ii}]/H$\alpha$ ratio, that are typical of ionized gas in a low metallicity dwarf galaxy (Boettcher et al. in preparation).
This stands in contrast to high forbidden to recombination line ratios in NGC~4418. Thus the ionized ISM observed in NGC~4418 does not consist of interstellar matter that was directly accreted from the interacting companion galaxy.
Thus while the interaction may have triggered changes in NGC~4418, we do not find evidence that direct gas exchanges are currently playing a role in the activities in the extended ionized-gas nebula.

One notable exceptional property about the dusty superwind in NGC~4418 is that both ionized and cold gases outflow from a nuclear component that is not at the systemic velocity of the host galaxy.
If we were to instead adopt the systemic velocity as an origin of the outflow, the velocity field on the NW side of the nucleus does not follow the monotonically blueshifting pattern with increasing distance from the nucleus, which is one of the key features of superwind in general.
Nonetheless, the superwind model seems the only feasible scenario that can naturally explain most other characteristics discussed so far (e.g., shock excitation, kpc-scale nebula size).
We found possible disk H~{\sc ii} regions in addition to the presumed superwind in this galaxy, but such H~{\sc ii} regions can neither explain the enhanced [N~{\sc ii}]/H$\alpha$ and [S~{\sc ii}]/H$\alpha$ ratios nor the velocity field that does not follow the galaxy disk rotation.
We note that any model of the nuclear activity in this galaxy needs to include this kinematically offset nuclear component from the host galaxy.
This is because the main activity of this galaxy is spatially and kinematically associated with this nuclear component, as clearly indicated in the velocity field along the major axis slit.
The emission lines are at the offset velocity from the systemic one only around the nuclear region, where the emission line fluxes peak (Section~\ref{results_emission_lines}).

\subsection{Other Observed Properties in the Context of Dusty Superwind Model}\label{dusty_superwind_detail}

The ``U'' shape in the optical reddening map is likely an envelope of the dusty superwind seen as a silhouette against the galaxy disk.
The enhanced extinction along the two vertical sidewalls of the ``U'' is probably due to projection effect of a hollow cone.
In order for the observers to see the silhouette of the polar dusty outflow on the NW side of the nucleus, this side of the disk corresponds to the far side.
We noted earlier that the stellar continuum is more extinct than the emission lines along the outflow (Section~\ref{results_cont}).
This can be explained if the dusty clouds, which either surround or are embedded within the cone-like ionized-gas nebula, efficiently extinct the stellar continuum in a screen configuration for the galaxy disk but only partly extinct the emission lines within the cone.

The signatures of the counter superwind to the opposite (SE) side of the nucleus, which is typically expected in superwind galaxies, are much less clear, although the nebula is also extended to this direction (Figure~\ref{minor_axis}).
The emission lines there do not show a monotonically redshifting pattern with increasing distance from the nucleus as expected for the outflow on the back side of the galaxy disk (Figure~\ref{minor_axis_k}).
This is probably because such outflow is obscured by dusty medium within the foreground disk.
In addition, the emission lines from disk H~{\sc ii} regions at $R_\mathrm{minor}\simeq -5''$ (Section~\ref{analysis_nebula}) further obscure the signatures.
However, our possible detection of Na~D emission around this region (Figure~\ref{focas_profmap}) might be such a counter wind.
Although Na~D emission is seldom observed in star forming regions, it is indeed observed on the back side of the dusty superwind in NGC~1808 \citep{phillips93}.

The enhanced emission-line ratios ([N~{\sc ii}]/H$\alpha\gtrsim 1$; [S~{\sc ii}]/H$\alpha\gtrsim 1$) seen almost everywhere in the extended ionized-gas nebula (Section~\ref{analysis_nebula}) are one of the key features for the identification of the superwind in NGC~4418.
Some other superwinds (e.g., ones in NGC~253 and NGC~1482; \citealt{veilleux02,westmoquette11}) also show similarly higher-than-unity line ratios.
On the other hand, the prototypical superwind in M82 shows less-enhanced ratios ([N~{\sc ii}]/H$\alpha=0.3$--0.6; e.g., \citealt{shopbell98}).
This is considered due to additional contribution of the stellar photoionization in the nuclear region \citep{shopbell98} and scattering of the nuclear emission lines by dusts within the superwind \citep{yoshida11}, both diluting the enhanced line ratios due to shock within the superwind.
In NGC~4418, the nuclear starburst is heavily obscured (see Section~\ref{discussion_keep_nucleus_dusty} for more discussion), and such dilution effect would not work.
The required shock velocity for [N~{\sc ii}]/H$\alpha\gtrsim 1$ and [S~{\sc ii}]/H$\alpha\gtrsim 1$ is 200--300 km~s$^{-1}$ according to the shock ionization model of \cite{allen08}.
Thd de-projected outflow velocity, on the other hand, is 190 km~s$^{-1}$ at most at the tip of the superwind (see below).
To enhance the line ratios than what are expected from the bulk outflow velocity, one needs an additional contribution of local velocity turbulence to the shock velocity.
The relatively broad (FWHM$\simeq 100$--200 km~s$^{-1}$) emission line width along the minor axis slit (Figure~\ref{minor_axis_k}) might indicate such velocity turbulence along the superwind.
We note, however, that such broad line width can also be explained by overlapping of two velocity components coming from near and far sides of the hollow outflowing cone if intrinsically double-peaked profile cannot be resolved with our resolution.

\subsection{Parameters Relevant to the Superwind Activity}\label{dusty_superwind_params}

The most important parameters that characterize the superwind were estimated with the following assumptions.
We assumed a simple bipolar outflow pointing toward the disk polar directions, as in the case of M82, although the superwind in NGC~4418 is more complicated than the one in M82 (Section~\ref{dusty_superwind_comp_detail} below).
We used the disk inclination angle to de-project the superwind geometry and the velocity field.
Total flux, mass, and kinetic energy of the superwind nebula were estimated by measuring them within the minor axis slit and scaling the quantities within this narrow ($0\farcs 8$ wide) slit to the typical width of the nebula ($4''$ wide) by assuming that the nebula is uniform along its width direction in velocity, emission-line flux, and column density.
This width corresponds to the size of the region that shows enhanced EWs of both H$\alpha$ emission and Na~D absorption ($\pm 2''$) along the major axis slit around the nucleus (Section~\ref{results_spatial_distribution}).
Although this central $\pm 2''$ area is much narrower than the total extent of the nebula along the offset-major axis slit ($\simeq 26''$ or 4.3~kpc; Figure~\ref{offset_major_axis}), it contains as much as 78\% of the total H$\alpha$ flux measured over the entire major axis slit.
We do not correct for extinction on the emission lines.
As for the total mass and kinetic energy of the superwind, we measured them only on the NW side of the nucleus and doubled the quantities by assuming bisymmetry to consider the counter superwind contribution.

We first estimated parameters of the ionized-gas superwind.
The de-projected ionized-gas superwind length from the nucleus is $\simeq 1.9$~kpc ($\simeq 10''$ in projection; Section~\ref{analysis_nebula}).
The total H$\alpha$ flux is $\simeq 1.8\times 10^{-14}$ erg~s$^{-1}$~cm$^{-2}$, and its luminosity is $\simeq 2.4\times 10^{39}$ erg~s$^{-1}$.
For comparison, \cite{lehnert95} reported H$\alpha$+[N~{\sc ii}] luminosity of $6.4\times 10^{39}$ erg~s$^{-1}$ (converted to our assumed distance) by using their narrow-band image.
We subtracted the [N~{\sc ii}] contribution by assuming it to be equal to the H$\alpha$ contribution (Section~\ref{results_emission_lines}), and found that their H$\alpha$ luminosity is $\simeq 30$\% larger than our measurement.
The total ionized-gas mass (including the assumed counter-superwind nebula) was estimated to be $\sim 1.2\times 10^5\ M_\mathrm{\sun}$ by assuming a case B condition for $T=10^4$ K and by using our measurements of electron density and volume filling factor (Section~\ref{analysis_nebula}).
Here, the mass is proportional to the total H$\alpha$ flux within the entire nebula divided by local electron density.
The uncertainty of the total flux is probably in the same order of the difference between our estimate and the narrow-band measurement by \cite{lehnert95}, i.e., $\sim 30$\%.
The local electron density is measured to be between the low density limit and 500~cm$^{-3}$ (Section~\ref{analysis_nebula}), and our adopted representative value (100~cm$^{-3}$) for this mass estimate likely has an uncertainty of about a factor of three.
The latter density uncertainty dominates the overall mass measurement uncertainty, and the derived mass above is uncertain by about a factor of three.
The de-projected outflow velocity at the NW tip of the superwind is $\simeq 190$ km~s$^{-1}$, and the dynamical age of the superwind is $\simeq 10$~Myr.
This age was estimated by dividing the superwind length by the outflow velocity at the tip, without considering the apparent velocity increase as a function of distance from the nucleus nor likely deceleration of the outflow velocity due to gravity of the host galaxy (see also the escape velocity argument below).
The uncertainty of the dynamical age is estimated to be $\sim 30$\% corresponding to the uncertainty of the inclination angle of $5^\circ$.
The kinetic energy of the ionized-gas superwind was estimated as $\frac{1}{2}\int_0^{r_{\rm out}} m(r) v(r)^2 dr$, where $m(r)dr$ is the mass of the ionized gas at the distance $r$ from the center within the superwind volume spanning small radial distance $dr$, $v(r)$ is the de-projected outflow velocity at position $r$, and $r_\mathrm{out}$ is the length of the superwind.
We found the total kinetic energy including the counter-superwind nebula contribution to be $\simeq 1.7\times 10^{52}$~erg.
Due to the uncertainty of the mass, the ionized-gas kinetic energy is also uncertain by about a factor of three.

We then estimated parameters of the cold-gas superwind.
The de-projected cold-gas superwind length from the nucleus is $\simeq 1.5$~kpc ($\simeq 8''$ in projection; Section~\ref{analysis_cold_gas}).
This length is a lower limit because detection of the absorption depends on the strength of the background stellar light.
The total cold-gas mass (including a counter-superwind nebula contribution) was estimated to be $\simeq 1.8\times 10^7\ M_\mathrm{\sun}$ by using the column density within the superwind.
This mass is proportional to the mean hydrogen column density within the superwind multiplied by the dusty superwind area.
We adopted the mean hydrogen column density from \cite{sakamoto13}, who used the optical extinction and the scaling to the hydrogen column density (Section~\ref{analysis_cold_gas}).
Their color uncertainty (about 0.1 mag, judging from the fact that they could differentiate extinctions at the peak and in its surrounding area having different extinction of 0.3 mag) corresponds to $\sim 10$\% of the column uncertainty.
The uncertainty of the dusty wind area is crudely estimated to be 50\%.
Therefore, the uncertainty of the cold gas mass is estimated to be about 50\% in total.
For comparison, \cite{sakamoto13} reported a total cold-gas mass of $4\times 10^7\ M_\mathrm{\sun}$ by using their optical color map.
The de-projected outflow velocity at the NW tip of the superwind is $\simeq 110$ km~s$^{-1}$.
The kinetic energy of the cold-gas superwind was estimated as $\frac{1}{2}\int_0^{r_{\rm out}} N_\mathrm{H} m_\mathrm{H} v(r)^2 w dr$, where $N_\mathrm{H}$ is the cold gas column density ($1.3\times 10^{21}$ cm$^{-2}$; Section~\ref{analysis_cold_gas}), $m_\mathrm{H}$ is the mass of a Hydrogen atom, $v(r)$ is the de-projected outflow velocity at position $r$, $w dr$ is the superwind surface area with width $w$ ($=4''$) and small radial distance $dr$, and $r_\mathrm{out}$ is the length of the superwind.
We found the total kinetic energy including a counter-superwind nebula contribution $\simeq 7\times 10^{53}$~erg.
This kinetic energy is about $\sim 40$ times larger than that of the ionized gas.
Due to the uncertainty of the mass, the cold-gas kinetic energy is also uncertain by about 50\%.

We finally estimated the escape velocity of the host galaxy.
In a simple gravitational model of truncated isothermal sphere, the local escape velocity at a distance $r$ from the nucleus is $V_\mathrm{esc.}=\sqrt{2}\times V_\mathrm{circ.}\times [1+\ln(r_\mathrm{max}/r)]^{0.5}$, where $V_\mathrm{circ.}$ is the disk rotation velocity, and $r_\mathrm{max}$ is truncation radius (e.g., \citealt{veilleux05}).
This formula has a very weak dependence on $r_\mathrm{max}/r$, and $V_\mathrm{esc.}=2.6$--$3.3\times V_\mathrm{circ.}$ at the range of distances of $r_\mathrm{max}/r=10$--100.
Although we have no good estimate of $r_\mathrm{max}$ for NGC~4418, the local escape velocity is very likely within $V_\mathrm{esc.}\simeq 390$--500 km~s$^{-1}$ for $V_\mathrm{circ.}=150$ km~s$^{-1}$ (Section~\ref{results_stellar_abs}).
This escape velocity is larger than the observed outflow velocity ($\simeq 190$ km~s$^{-1}$) and, thus, the off-planar material transported with the superwind will not escape the host galaxy potential and will eventually fall back.

\subsection{Comparison with M82}\label{dusty_superwind_comp_M82}

\subsubsection{Comparison of basic characteristics}

It is helpful to compare NGC~4418 with M82, the prototypical starburst galaxy with superwind, in more detail to further characterize the superwind in NGC~4418.
We adopted a distance to M82 of 3.25~Mpc \citep{tammann68}.
Table~\ref{table_comp_n4418_m82} summarizes various quantities of the two galaxies and their superwinds.

Regarding their host galaxies, NGC~4418 is an early-type spiral (Sa) galaxy.
M82, on the other hand, is usually classified as irregular, but is more likely a late-type barred spiral (SBc) galaxy \citep{mayya05}.
Their disk rotation velocities and dynamical masses are roughly comparable to each other \citep{greve11,greco12}.
NGC~4418 is 0.4 mag fainter than M82 in absolute $K$-band magnitude \citep{2mass}.
Regarding their nuclear activities, the infrared luminosity ($L_\mathrm{IR}$) of NGC~4418 is three times more \citep{sanders03}, but its 1.4~GHz luminosity is only about half \citep{klein88,yun01,condon02} of M82.
Together, the FIR-to-radio luminosity ratio is $\sim 5$ times higher in NGC~4418.

Regarding their superwind properties, we compare the superwind parameters in NGC~4418 with those of the filamentary component of the nebula in M82.
\cite{shopbell98} identified the halo component in addition to the filamentary one, and attributed the halo to the nuclear and/or disk contributions.
The ionized-gas mass within the superwind is much larger by $\sim 50$ times in M82 \citep{lehnert95,shopbell98}.
The superwind length is about twice larger, and the outflow velocity is about a few times faster, in M82 \citep{shopbell98}.
Together, the kinetic energy of the ionized-gas superwind in NGC~4418 is three orders of magnitude smaller than in M82, and the dynamical age is longer by a factor of a few in NGC~4418 \citep{bland88,shopbell98}.
The cold-gas column density within the superwind is comparable to each other, but mass of the cold gas is larger by a factor of $\sim 2$ in M82 \citep{heckman90}.
The kinetic energy of the cold-gas superwind in M82 is not well known, but that in NGC~4418 is about one tenth of the ionized-gas kinetic energy in M82.

\begin{table*}[ht]
\caption{Comparisons of starburst and superwind parameters between NGC~4418 and M82.\label{table_comp_n4418_m82}}
\scriptsize
\begin{tabular}{clcccc}
\tableline\tableline
\multicolumn{2}{l}{Parameters (unit)} & NGC~4418 & references & M82\tablenotemark{a} & references \\
\tableline
\multicolumn{6}{l}{Host galaxy} \\
(1) & Morphological type & (R')SAB(s)a & \cite{rc3} & I0 edge-on & \cite{rc3} \\
 &                                &      &                & SBc & \cite{mayya05} \\
(2) & $M_\mathrm{K}$ (mag) & $-22.5$ & \cite{2mass} & $-22.9$ & \cite{2mass} \\
(3) & $V_\mathrm{circ.}$ (km~s$^{-1}$) & $\simeq 150$ & this work & 130 & \cite{greve11,greco12} \\
(4) & $M_\mathrm{dyn}$ ($M_\mathrm{\sun}$) & $\sim 0.8\times 10^{10}$ & this work & $1.0\times 10^{10}$ & \cite{greco12} \\
\multicolumn{6}{l}{} \\
\multicolumn{6}{l}{Nucleus activity} \\
(5) & $L_\mathrm{IR}\ (L_\mathrm{\sun})$ & $1.4\times 10^{11}$ & \cite{sanders03} & $4.7\times 10^{10}$ & \cite{sanders03} \\
(6) & $L_\mathrm{1.4~GHz}$ (W Hz$^{-1}$) & $4.2\times 10^{21}$ & \cite{yun01} & $1.0\times 10^{22}$ & \cite{klein88} \\
 &                                                    & $5.6\times 10^{21}$ & \cite{condon02} & & \\
\multicolumn{6}{l}{} \\
\multicolumn{6}{l}{Ionized-gas superwind} \\
(7) & $L_\mathrm{H\alpha}$ (erg~s$^{-1}$) & $2.4\times 10^{39}$ \tablenotemark{b} & this work & $2\times 10^{41}$ (total)\tablenotemark{c} & \cite{heckman90} \\
 & & $3.2\times 10^{39}$ \tablenotemark{b,d} & \cite{lehnert95} & $7.6\times 10^{40}$ (filaments) & \cite{shopbell98} \\
(8) & Mass ($M_\mathrm{\odot}$) & $1.2\times 10^5$ \tablenotemark{e,g} & this work & $5.8\times 10^6$ (filaments) & \cite{shopbell98} \\
(9) & Length (kpc) & 1.9 & this work & 3.5 & \cite{shopbell98} \\
(10) & Velocity (km~s$^{-1}$) & 190 \tablenotemark{f} & this work & 525 (inner); 655 (outer) & \cite{shopbell98} \\
(11) & Kinetic energy (erg) & $1.7\times 10^{52}$ \tablenotemark{e,g} & this work & $2.1\times 10^{55}$ (filaments) & \cite{shopbell98} \\
(12) & Dynamical age (Myr) & 10 \tablenotemark{h}& this work & 3 & \cite{shopbell98}\tablenotemark{h} \\
\multicolumn{6}{l}{} \\
\multicolumn{6}{l}{Cold-gas superwind} \\
(13) & $N_\mathrm{H}$ (cm$^{-2}$) & 0.3--$10\times 10^{21}$ & this work & $1.3\times 10^{21}$ & \cite{heckman90} \\
 &                                               & $1.3\times 10^{21}$ & \cite{sakamoto13} & & \\
(14) & Mass ($M_\mathrm{\sun}$) & $1.8\times 10^7$ \tablenotemark{e,i} & this work & $6\times 10^7$ & \cite{heckman90} \\
 &                                                   & $4\times 10^7$ & \cite{sakamoto13} & & \\
\tableline
\end{tabular}
\tablenotetext{a}{Converted to a distance of 3.25~Mpc \citep{tammann68}.}
\tablenotetext{b}{Uncorrected for the internal extinction.}
\tablenotetext{c}{Corrected for the internal extinction.}
\tablenotetext{d}{Contribution of [N~{\sc ii}] is subtracted by assuming [N~{\sc ii}]/H$\alpha=1$. Converted to our assumed distance.}
\tablenotetext{e}{The counter-wind contribution is considered (see text).}
\tablenotetext{f}{Velocity at the tip of the outflow.}
\tablenotetext{g}{With uncertainty of factor three (see text).}
\tablenotetext{h}{With uncertainty of $\sim 30$\% (see text).}
\tablenotetext{i}{With uncertainty of $\sim 50$\% (see text).}
\tablecomments{
(1) morphological type, (2) $K$-band absolute magnitude (Vega) of the entire galaxy, (3) de-projected disk rotation velocity, (4) dynamical mass from the stellar rotation velocity, (5) infrared luminosity at 8--$1000\ \mu$m, (6) 1.4~GHz luminosity of the nucleus, (7) H$\alpha$ luminosity of the superwind, (8) ionized-gas mass, (9) de-projected superwind length, (10) de-projected outflow velocity, (11) kinetic energy of the ionized-gas outflow, (12) dynamical age of the superwind, (13) hydrogen column density within the superwind, and (14) cold-gas mass within the superwind.}
\end{table*}

\subsubsection{More complicated superwind in NGC~4418}\label{dusty_superwind_comp_detail}

The superwind in NGC~4418 seems to show more complicated characteristics when compared to the one in M82 despite the overall qualitative similarities between the two.
The nebula in NGC~4418 apparently extends with a large opening angle from the nucleus, and it is not clear if the tangentially extended part of the nebula is also associated with the superwind found along the disk semiminor axis.
Along the offset-major axis slit, the nebula tangentially spans over $\sim 3.3$~kpc ($\sim 20''$) (Figure~\ref{offset_major_axis}).
Interestingly, it extends to well outside of the ``U''-shaped region that probably outlines the dusty superwind outflow.
Judging from both kinematics and physical conditions of the ionized gas, the widely extended nebula is most likely to be a continuation of the superwind along its outflow axis.
The emission line ratios in this extended region are more enhanced ([N~{\sc ii}]/H$\alpha\sim 1$; [S~{\sc ii}]/H$\alpha\sim 1$) than those in typical H~{\sc ii} regions, similar to the ratios near the outflow axis within the ``U'' along the disk semiminor axis.
The velocity is mostly blueshifted and changes smoothly along the offset-major axis slit, including regions near the disk semiminor axis.
Such velocity field is quite different from that of stars in the disk, but it connects to that of the outflowing ionized gas along the disk semiminor axis at the intersection of the slits (Figure~\ref{offset_major_axis_k}).
These characteristics look quite different from the halo component of the superwind in M82.
This halo in M82 looks smoothly and exponentially extended and apparently surrounds filamentary conical superwind, and it is most likely due to scattering of the nuclear starburst (and also disk) emission by dusts within the galaxy halo (e.g., \citealt{bland88,shopbell98}).
In NGC~4418, we found an extended nebula that is also wide along the major axis of the disk outside of the bottom part of the ``U'' ($2''\lesssim |R_\mathrm{major}|\lesssim 10''$).
The extended nebula is not due to disk H~{\sc ii} regions, because its emission-line ratios are enhanced ([N~{\sc ii}]/H$\alpha\sim 1$; [S~{\sc ii}]/H$\alpha\sim 1$) and its velocity field does not follow the galaxy disk rotation (Figures~\ref{offset_major_axis},~\ref{offset_major_axis_k}).
The outflowed material is found to eventually fall back to the galaxy disk by using the escape velocity argument (Section~\ref{dusty_superwind_params}).
The originally ``U''-shaped outflow might extend wider at higher galactic latitudes before falling down to the disk region, like a fountain.
The falling-back material can be shock heated and this may explain the enhanced line ratios.
We note, however, that, because the age of the superwind-driving starburst and the dynamical age of the superwind is similar (see Section~\ref{consistency_starburst_superwind} below), it seems likely that such fountain has not yet developed well in NGC~4418.
It is known that M82 also shows similarly enhanced line ratios in its disk, and \cite{shopbell98} suggested a possibility of mixing-layer heating due to hot X-ray emitting plasma associated with the superwind activity.
This may also be the reason for the enhanced emission-line ratios in NGC~4418.
We need more detailed observations to understand the types of activities ongoing along the disk of NGC~4418.

The NGC~4418 superwind is non-axisymmetric about the disk polar axis, in contrast to the almost axisymmetric prototypical superwind in M82.
First, the NE half of the superwind is more extinct, especially near the nucleus (Section~\ref{results_optical_image} and Figure~\ref{findingchart}).
Second, the ionized-gas velocity field along the offset-major axis slit is not symmetric about the disk semiminor axis, showing more blueshift on the SW side of the semiminor axis (Figure~\ref{offset_major_axis_k}).
This velocity trend is opposite from that along the major axis showing more blueshift on the NE side of the seminar axis (Figure~\ref{major_axis_k}), being inconsistent with the rotating outflow model \citep{heckathorn72,sofue92,veilleux94,seaquist01,veilleux02,walter02,greve04}.
An alternative model to explain such velocity non-axisymmetry is an outflow tilted from the disk polar direction.
We note, however, that the emission line fluxes are distributed almost symmetrically about the disk semiminor axis along the offset-major axis slit (Figure~\ref{offset_major_axis}), suggesting instead that the outflow axis does not show a large tilt.

Such non-axisymmetric superwind morphology and the velocity field can be caused by non-uniform collimation near the base of the outflow.
The superwind in M82 is often modeled with a simple hollow cone pointing toward the disk polar directions (e.g., \citealt{mckeith95,walter02,greve04}; but see also \citealt{shopbell98} for detailed descriptions of its slight asymmetry).
Such structure is naturally expected for nuclear starburst in disk galaxies, in which hot gas heated by numerous SNe expands preferentially toward the disk polar directions due to collimation by dense gas in the disk (e.g., \citealt{heckman90}).
In NGC~4418, \cite{sakamoto13} found that most molecular gas is concentrated at the nucleus at the scale of 100~pc, where the optical extinction is largest (Section~\ref{results_optical_image}) and the superwind-driving activity is likely located (see Section~\ref{discussion_where_is_starburst} below).
In addition, they found molecular gas extended toward north from the concentration at the scale of $<3''$, where we also found the extended dusty region along bottom of the ``U'' (Section~\ref{results_optical_image}).
Such association of the dusty regions to the molecular gas has been already pointed out by \cite{sakamoto13}, and our study made this point clearer.
\cite{sakamoto13} argued that the wind might have initially blown off toward north from the nucleus, and changed the direction toward NW due to interaction with the molecular gas there.
Although we could not test this idea without our slit coverage there, it seems very likely that such non-uniform molecular gas distribution near the galaxy nucleus plays a role to shape the asymmetry of the dusty superwind at the larger scale.
Alternatively, it is conceivable that the initially symmetric wind is disturbed by a high-latitude gas that has been provided as a result of the interaction with a neighbor galaxy VV~655 \citep{varenius17} and is likely distributed non-uniformly around NGC~4418.

\section{DISCUSSION}

\subsection{Starburst--Superwind Scenario}\label{discussion_starburst_superwind_scenario}

We have found that the dusty superwind model best describes the observed characteristics of both ionized and cold gases off the nucleus of NGC~4418.
We also found that the nucleus is composed of a starburst population by using the spectral synthesis analysis (Section~\ref{results_spectral_synthesis}).
Here a question arises whether this nuclear starburst is the driver of the superwind, and if so whether the starburst--superwind scenario is consistent with other observed characteristics.

\subsubsection{Can the nuclear starburst drive the superwind?}\label{consistency_starburst_superwind}

\begin{figure*}
\plotone{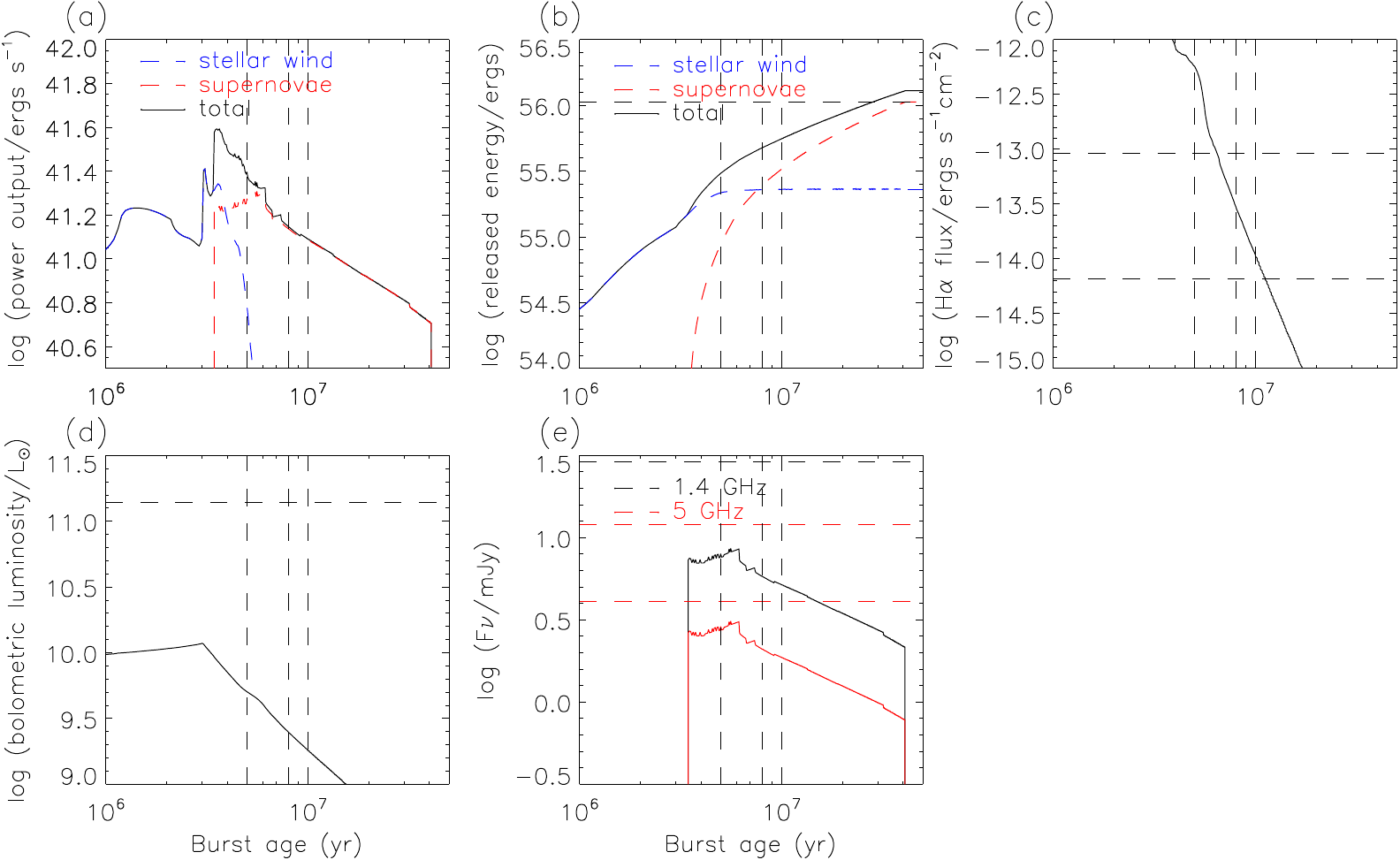}
\caption{
\explain{We reset the plotting range in Panel (c) so that the model value at 5~Myr can be nicely shown.}
Evolution of various starburst and superwind-related properties as a function of the burst age.
(a) power output from both stellar wind (blue), SNe (red), and sum of the two (black), (b) released energy from the respective components in (a), (c) (extinction-free) H$\alpha$ emission-line flux, (d) bolometric luminosity, and (e) radio fluxes at 1.4~GHz (black) and 5~GHz (red).
In each panel, relevant observed quantities are marked with horizontal broken lines.
(b) required energy to drive the cold-gas superwind (the efficiency of 1\% is assumed; see text), (c) observed (lower) and extinction-corrected (upper; by using $A_\mathrm{V}$ inferred from the spectral synthesis analysis) SDSS H$\alpha$ emission-line fluxes, (d) infrared luminosity ($L_\mathrm{IR}$), and (e) radio fluxes from the extended component at both 1.4~GHz (black) and 5~GHz (red; there are two measurements; see text).
The fiducial burst age (10~Myr) and other ages discussed in our analysis (5~Myr and 8~Myr; Section~\ref{results_spectral_synthesis}) are marked with vertical broken lines in each panel.
\label{starburst99}}
\end{figure*}

We first examined if the dynamical age of the superwind is consistent with the starburst age derived from the spectral synthesis, and if the kinetic energy of the superwind can be supplied from the starburst.
We modeled the nuclear starburst with {\it starburst99} (version 7.01; \citealt{starburst99}; see also \citealt{starburst05,starburst14}) and found that this starburst can develop the superwind as observed.
In the {\it starburst99} model, we adopted the same parameters as for the spectral synthesis, namely, solar-metalicity instantaneous burst with the Padova stellar evolutionary isochrones with thermally pulsing AGB stars.
We also adopted the stellar mass of the starburst derived by the spectral synthesis ($\simeq 1\times 10^7\ M_\mathrm{\sun}$ for the burst age of 8~Myr; Section~\ref{results_spectral_synthesis}), and a Kroupa universal IMF whose upper and lower mass cutoffs are $100\ M_\mathrm{\sun}$ and $0.1\ M_\mathrm{\sun}$, respectively.
Figure~\ref{starburst99}a shows evolution of power output from the starburst through both stellar wind and SN explosions.
The stellar wind from massive stars dominates the output at $\lesssim 4$~Myr, when SNe start to provide kinetic energy into the circumnuclear region.
After a short period when both stellar wind and SNe contribute to the power, SNe dominate the output at $\gtrsim 5$~Myr.
Together, the total power remains in the same order up to $\sim 10$~Myr, although this result depends on the adopted IMF at $>8\ M_\mathrm{\sun}$ and the fate of very massive stars either as core-collapsed SNe or failed SNe \citep{spera15}.
Given such a power output evolution, the starburst--superwind scenario requires that the dynamical age of the superwind is close to the starburst age.
We indeed found the superwind dynamical age ($\simeq 10$~Myr; Section~\ref{dusty_superwind_params}) is very similar to the starburst age (8--10~Myr; Sections~\ref{results_spectral_synthesis},~\ref{scenario_consistency_ha_ir}), i.e., the starburst--superwind scenario satisfies the time scale requirement.
In the following, we adopted 10~Myr as the starburst age.
Figure~\ref{starburst99}b shows evolution of the total released energy from stellar wind and SNe; this is the time integration of the power in Figure~\ref{starburst99}a.
We found that the expected released energy of this starburst at 10~Myr is $\simeq 6\times 10^{55}$~erg.
This is very similar to the requirement to drive the cold-gas superwind outflow, whose kinetic energy is $\simeq 7\times 10^{55}$~erg.
Here, we assumed a conversion efficiency of 1\% from the energy released from the starburst to the kinetic energy of the cold-gas outflow \citep{martin06}.
Given considerable uncertainties in the measurements of the kinetic energy and the conversion efficiency, the starburst--superwind scenario satisfies the energy budget requirement.

\subsubsection{Consistencies of the starburst--superwind scenario in H$\alpha$ flux and infrared luminosity}\label{scenario_consistency_ha_ir}

We then examined consistency of the H$\alpha$ flux and the infrared luminosity between the starburst model and the observations.
Figure~\ref{starburst99}c shows evolution of (extinction-free) H$\alpha$ flux, and we found $1.0\times 10^{-14}$ erg~s$^{-1}$~cm$^{-2}$ at 10~Myr.
Because the nuclear ionized gas is predominantly shock-excited (Section~\ref{analysis_nebula}) and not powered by this starburst, the starburst prediction should not dominate the observed flux.
The extinction-corrected (for $A_\mathrm{V}=3.5$ mag as inferred from the Balmer decrement; Section~\ref{results_spectral_synthesis}) SDSS H$\alpha$ flux is $\simeq 9\times 10^{-14}$ erg~s$^{-1}$~cm$^{-2}$, being a factor $\sim 9$ larger than the starburst prediction as required by the scenario.
The younger 5~Myr starburst, which is as young as the inferred age of the core cluster, emits much more H$\alpha$ luminosity than the observational limit.
Number of the ionizing photons rapidly increases for younger starburst without changing the spectral slope very much in the blue spectral region (Figure~\ref{starburst99}c; e.g., \citealt{starburst99}).
We found that the 5~Myr starburst emits $\simeq 30$ times more H$\alpha$ luminosity than the 10~Myr starburst, giving $\sim 3$ times more H$\alpha$ luminosity than the extinction-corrected observed one that includes a contribution from the shock.
Here, we considered only difference in mass-to-luminosity ratio for starbursts with different ages in the same way as in Section~\ref{results_spectral_synthesis}.

Figure~\ref{starburst99}d shows evolution of the bolometric luminosity of the starburst, which rapidly decreases after 3~Myr, and we found $2\times 10^9\ L_\mathrm{\sun}$ at 10~Myr.
This luminosity corresponds to $\sim 1$\% of the current infrared luminosity ($L_\mathrm{IR}$) of the galaxy and, therefore, this 10~Myr-old starburst can be a part of NGC~4418.

\subsubsection{Consistency of the starburst--superwind scenario in radio fluxes}\label{consistency_radio}

We further examined consistency of the radio continuum fluxes between the starburst model and the observations.
We first summarize here the radio observations at both 1.4~GHz and 5~GHz.
At 1.4~GHz, the nuclear radio source seems to consist of both compact and extended sources.
\cite{condon90} found that the radio emission from this galaxy at 1.49~GHz is confined to the nucleus at the scale of $500\times 300$ milli-arcsecond (mas), because observations at various lower resolutions show very similar fluxes ($\simeq 40$~mJy).
\cite{costagliola13} found a smooth extended (FWHM$\simeq 500$ mas) component (29~mJy; hereafter, the extended component) in addition to an unresolved source with a smaller beam of $350\times 160$ mas at 1.41~GHz (38~mJy in total).
At 5~GHz, the nuclear radio source seems more compact than the extended 1.4~GHz component.
\cite{baan06} found an almost unresolved source (26.1~mJy) with a beam of $180\times 20$ mas at 4.8~GHz.
\cite{costagliola13} found a slightly extended component (FWHM$\simeq 150$ mas) and an unresolved source with a beam of $130\times 40$ mas (34~mJy in total) at 5~GHz.
We note that their 5~GHz image (their Figure~1) seems to show an even more extended (FWHM$\simeq 500$ mas) component.
\cite{varenius14} resolved the compact radio source into eight individually very compact ($<49$ mas) blobs (22~mJy in total) distributing within 250 mas with a beam of $21\times 15$ mas at 5.0~GHz.
They identified this collection of blobs (hereafter, the compact component) with the sub-mm core found by \cite{sakamoto13}.
This compact component seems to correspond to the unresolved and slightly extended (FWHM$\simeq 150$ mas) components of \cite{costagliola13}.
We then estimated the 5~GHz flux of the extended component by subtracting the flux of the compact component (22~mJy; \citealt{varenius14}) from the total flux, and found 4.1~mJy ($26.1-22$~mJy) or 12~mJy ($34-22$~mJy) by adopting either the total flux measurements of \cite{baan06} or \cite{costagliola13}, respectively.

The radio spectral index ($\alpha$, where $S_\nu\propto\nu^\alpha$) between 1.4~GHz and 5~GHz shows a clear contrast between the two components.
It is inverted ($\alpha\gtrsim 0.7$; \citealt{varenius14}) for the compact source and steep ($\alpha=-1.6$ or $-0.7$ for the total fluxes of \cite{baan06} or \cite{costagliola13}, respectively) for the extended component.
\cite{varenius14} investigated the origin of this unusual inverted index of the compact component, and discussed possibilities of synchrotron self-absorption and thermal free-free absorption.
On the other hand, the steep spectral index of the extended component is consistent with that of non-thermal emission ($\alpha\simeq -0.8$; \citealt{condon92}).

We then examined if the starburst model is consistent with these observations, and found that it is consistent with the extended component.
Figure~\ref{starburst99}e shows evolution of the non-thermal radio fluxes at both 1.4~GHz and 5~GHz of our starburst model.
Here, {\it starburst99} is used to predict the rate of SN explosion, and we converted it to the radio flux in the figure using equation (17) of \cite{bressan02} with a radio spectral index of $\alpha=-0.8$ and neglecting the radio SN contribution ($\simeq 5$\% of the total non-thermal flux).
We replaced the average luminosity per SN event in the original equation of \cite{bressan02} with a recent recalibration by \cite{obi17}, giving 35\% more fluxes in both bands than the original calibration.
We found 5.2~mJy and 1.8~mJy at 10~Myr, corresponding to $1/6$ and $1/7$--$1/2$ of the observed fluxes at 1.4~GHz and 5~GHz, respectively.
Therefore, the starburst can be a part of NGC~4418.
As we noted above, this starburst model underpredicts the energy output required to drive the superwind by about a factor of two, showing the similar sense in the radio flux underprediction.
This might suggest that the stellar mass of the starburst needs to be doubled for better match.
We found no problems in explaining the steep radio spectral index in the radio observations, because radio emission is predominantly non-thermal at 10~Myr from many SNe already explored by that time.
Meanwhile, as for the compact component, this starburst model can account for only $\lesssim 10$\% of the observed 5~GHz flux, indicating that one needs an additional component that shows unusual radio spectral index in the compact region.
This seems not surprising because our starburst model is constructed in the optical wavelengths, where the sub-mm core is highly obscured (Section~\ref{results_spectral_synthesis}).
It is therefore likely that the sub-mm core has most of the 5~GHz flux of the compact component.
We note that we cannot use the observed core flux at 1.4~GHz for the comparison with the model, because the observed inverted radio spectral index indicates optically thick emission at this frequency \citep{varenius14}.

\subsection{How Does the Nucleus Remain Embedded in Dust?}\label{discussion_keep_nucleus_dusty}

How does the nuclear region of NGC~4418 remain embedded in dust after experiencing the superwind that expelled the dusty circumnuclear material observed as the superwind?
We found that the mass outflow rate estimated in this work is roughly comparable to the mass inflow rate measured in earlier studies, suggesting that the dusty nuclear environment can be conserved by the supply of the cold dusty material.
The cold gas removal rate from the nucleus was estimated to be $\sim 2\ M_\mathrm{\sun}$~yr$^{-1}$ by using the mass of the cold gas within the superwind ($\simeq 1.8\times 10^7\ M_\mathrm{\sun}$) and the superwind dynamical age ($\simeq 10$~Myr; Section~\ref{dusty_superwind_params}).
As noted earlier such gas cannot leave the galaxy (Section~\ref{dusty_superwind_params}) and, therefore, this outflow rate refers to the rate at which the gas leaves the nuclear region.
The mass inflow rate toward the nucleus has been estimated with redshifted absorptions of the low-lying OH doublets and [O~{\sc i}] $63\ \mu$m in FIR ($\lesssim 12\ M_\mathrm{\sun}$~yr$^{-1}$; \citealt{ga12}) and redshifted H~{\sc i} emission (11--$49\ M_\mathrm{\sun}$~yr$^{-1}$; \citealt{costagliola13}).
We caution that these inflow rates are based on the line-of-sight absorption measurements and assume an isotropic inflow.
Because we established an outflow motion in this work, these inflow rates are overestimated and, thus, upper limits.
\cite{ga12} and \cite{costagliola13} argued that the inflow is likely taking place at an inner $\sim 100$~pc region on the basis of their detailed multi-transition molecular-line analysis.
We note that \cite{veilleux13} showed by using very similar OH line spectroscopy that luminous galaxies typically show molecular outflows, and NGC~4418 is a rare exception to show an inflow.
They argued that such an inflow is a short timescale phenomenon in a compact region, or that the inflowing gas subtends a relatively small fraction of $4\pi$ steradians.
In the latter scenario, the inflow takes place preferentially in the disk plane, a part of which is observed in absorption along our line-of-sight toward the compact nucleus, whereas the outflow takes place preferentially toward the disk polar directions.
\cite{sakamoto13} suggested the latter possibility for the case of NGC~4418 on the basis of their finding of the dusty superwind, and our study added more support to this.
We note, however, that the circumnuclear region is likely kinematically disturbed (Section~\ref{co_counterrotate}), and the actual inflow/outflow geometry is probably very complex.
Such disturbed kinematics might help the dusty circumnuclear gas to loose its angular momentum, enabling efficient inflow toward the very central region of this galaxy.

\subsection{The Core in the Context of the Starburst--Superwind Scenario}

\subsubsection{Where are the superwind-driving starburst and the sub-mm core?}\label{discussion_where_is_starburst}

Both the sub-mm core and the superwind-driving starburst are likely embedded in the circumnuclear dusty molecular gas concentration in the central $<1''$, which is typical spatial resolution in the optical analysis in this work.
The molecular gas concentration corresponds to the area of large optical extinction in the nucleus, because the circumnuclear dusty region at the scale of a few arcsec is very similar in morphology to the circumnuclear molecular-gas distribution (Section~\ref{dusty_superwind_comp_detail}).
The nebula emission-line fluxes peak at the peak of the optical extinction (Sections~\ref{results_optical_image},~\ref{results_emission_lines}), implying that the starburst is there.
The sub-mm core is also located in the molecular gas concentration \citep{sakamoto13}.
Therefore, the starburst is in the vicinity of the core.
However, the core cluster is unlikely to have driven the superwind.
This is for the following reasons.
First, the burst age of this core cluster inferred in the sub-mm study is $\lesssim 5$~Myr \citep{sakamoto13}, being inconsistent with the age of the superwind-driving starburst ($\simeq 8$--10~Myr).
We argued that the 5~Myr starburst would produce too much H$\alpha$ luminosity when compared to the observation (Section~\ref{scenario_consistency_ha_ir}), i.e., the superwind-driving starburst cannot be as young as the core cluster.
Second, the mass of the core cluster ($10^{8.0\pm 0.5}\ M_\mathrm{\sun}$ ($r\lesssim 10$~pc); \citealt{sakamoto13}) is larger than the stellar mass of the superwind-driving starburst (1.0--$1.4\times 10^7\ M_\mathrm{\sun}$ for the burst age of 8--10~Myr; Section~\ref{results_spectral_synthesis}).
We note that the younger 5~Myr starburst requires smaller stellar mass due to smaller mass-to-light ratio, making the mass discrepancy larger between the two.
Third, the superwind-driving starburst likely contributes only a tiny fraction ($\sim 1$\%) of the observed infrared emission, although it is powerful enough to drive the superwind.
This is in contrast to the sub-mm core having a significant fraction of the total bolometric luminosity of the galaxy.
Fourth, a signature of the starburst population is detected in the optical spectrum (Section~\ref{results_spectral_synthesis}), whereas the core, either the core cluster or an AGN, is almost completely obscured at near-infrared and shorter wavelengths.
The high-resolution {\it HST} image at $\simeq 2\ \mu$m does not show any compact structure corresponding to the sub-mm core ($\sim 20$~pc; \citealt{scoville00,evans03}).
This last reason is independent of our model-dependent analysis results based on spectral synthesis and starburst evolutionary modeling.
In this way the nuclear region of NGC~4418 contains both the core and the superwind-driving starburst.
Due to their proximity to each other, the feedback processes are/have been likely affecting their formations and the activities (Section~\ref{discussion_con_and_superwind}), although our optical observations did not show their direct hints in the circumnuclear region.
There remains a possibility that the most inner part of the outflow in the vicinity of the core is driven by the core behind the enshrouding dusts, because the core is more massive and providing more energy output than the superwind-driving starburst.
We note that composite of a compact luminous core, circumnuclear starburst, and an outflow of gas and dust is similar to the western nucleus of Arp~220 \citep{sakamoto08,sakamoto17}.

What are the counterparts of the superwind-driving starburst among various known circumnuclear components studied earlier in other wavelengths?
Given the presence of a massive luminous infrared emitter in addition to the starburst, many of the known distinct CON characteristics of this nucleus (e.g., presence of a distinct MIR core, large FIR-to-radio luminosity ratio; see Section~\ref{intro} for the detailed descriptions) are likely due to, in significant part, the core activities rather than the starburst.
Therefore, identifying the starburst-related activities is important to understand the activities of both the CON and the starburst itself.
\cite{costagliola13} found a redshifted ($\delta V\simeq 90$ km~s$^{-1}$) and broad (FWHM$\simeq 160$ km~s$^{-1}$) CO (2--1) emission, covering the redshift of the starburst, almost at the same positions of the systemic component and the mm continuum emission.
They estimated a mass of 1--$4.2\times 10^7\ M_\mathrm{\sun}$ for this redshifted component, which is about one fifth of the systemic-velocity component mass and in the same order of the stellar mass of the starburst inferred from our spectral synthesis.
\cite{sakamoto13} also found similarly broad ($\simeq 350$ km~s$^{-1}$ full width near zero flux level) and redward-asymmetric (up to $\delta V \simeq +250$ km~s$^{-1}$) CO (3--2) emission at the nucleus, covering the redshift of the starburst.
Unlike \cite{costagliola13}, \cite{sakamoto13} interpreted this redward-asymmetric profile as due to redshifted self absorption.
Regardless of the interpretation of the CO profile, it seems very likely that significant amount of the circumnuclear molecular gas exists where the starburst likely exists in terms of both the position and the velocity.
\cite{costagliola13} also found an extended redshifted H~{\sc i} absorption against the extended radio continuum emission.
According to their model of triple-layer spherical shells around the core, the H~{\sc i} absorption is due to the infalling colder envelope at the scale of $\sim 100$~pc.
As noted earlier (Section~\ref{consistency_radio}), the starburst can be the extended (500 mas, or $\sim 80$~pc) radio component because of the consistency in the radio fluxes and the radio spectral index between the starburst model and the observations.
If this is the case, this extended starburst provides the background continuum on which the H~{\sc i} absorption of the surrounding and infalling colder material is detected.

We here briefly summarize nuclear components of NGC~4418, their roles and configuration within the nucleus that have been discussed so far in this paper and studied earlier in the literature.
The $\simeq 10$~Myr-old superwind-driving starburst is located at the nucleus but at $\simeq 40$ km~s$^{-1}$ off the host galaxy systemic velocity (Sections~\ref{systemic_velocity},~\ref{dusty_superwind_evidence}).
It is located behind the nuclear dust lane in the optical wavelengths (Sections~\ref{results_optical_image},~\ref{discussion_where_is_starburst}) but is detected in radio as a 500 mas ($\sim 80$~pc) component particularly at lower frequency of 1.4~GHz because of the steep spectral index (Section~\ref{consistency_radio}).
The cold inflowing H~{\sc i} gas is distributed in front of this starburst, causing redshifted H~{\sc i} 21~cm absorption seen against this extended continuum source \citep{costagliola13}.
The very compact ($<20$~pc) sub-mm core is located in the nuclear region \citep{sakamoto13} in addition to the starburst.
Even if the sub-mm core contains a starburst, its properties inferred in the sub-mm study are incompatible with those of the superwind-driving starburst (Section~\ref{discussion_where_is_starburst}).
The core is buried in the NIR and shorter wavelengths \citep{scoville00,evans03}, but emits most of the infrared emission of this galaxy (Section~\ref{scenario_consistency_ha_ir}).
The strong MIR emission with very deep silicate absorption around $10~\mu$m \citep{spoon01,evans03,siebenmorgen08} is likely a part of this infrared emission from the core.
The sub-mm core is composed of very compact 10~pc-scale blobs, and the blobs collectively show inverted radio spectral index due to either synchrotron self-absorption or thermal free-free absorption \citep{varenius14}.
\cite{varenius14} argued that a single AGN cannot power the entire core.

\subsubsection{New constraint on the core activity in the nucleus hosting the superwind}\label{discussion_con_and_superwind}

Our finding of the recent superwind activity off the core puts a unique constraint on the timing of activation of the core and formation timescale of the massive molecular gas concentration around the core.
For the superwind-driving starburst to occur $\simeq 10$~Myr ago, one needs a massive molecular gas concentration in the nuclear region, from which the stars formed.
If the core was luminous and not shielded by the dusty shell at that time, it was probably difficult for the starburst to happen due to harsh environment for the cold molecular gas to exist because of, e.g., strong radiation pressure and/or ionization by intense UV radiation from the naked core.
In addition, it seems very difficult to later enshroud the bare luminous core to make it as we see it today, because the shroud should see direct influence of the activity of the core.
On the other hand, if the luminous core was enshrouded then, the dusty envelope itself should have been easily removed by now because of the similar reasons.
Therefore, the core likely became luminous in the dusty shroud after the superwind-driving starburst occurred.
Note that this superwind outflow should have removed a significant fraction of the dusty nuclear material.
About one third of the present-day nuclear molecular gas concentration ($\sim 10^{8.0}\ M_\mathrm{\sun}$ at $r\lesssim 10$~pc; \citealt{sakamoto13}) can be removed within the dynamical age of the superwind ($\simeq 10$~Myr) at a rate of $\sim 2\ M_\mathrm{\sun}$~yr$^{-1}$.
Although the cold material inflow can keep the circumnuclear region dusty (Section~\ref{discussion_keep_nucleus_dusty}), it is not clear if the massive and compact molecular gas concentration surrounding the core can be formed after experiencing the superwind activity only $\simeq 10$~Myr ago.
To generate present-day molecular gas concentration after the cleaning-up by the superwind, one needs very rapid mass inflow toward the very central part of the nucleus.
One may also need a spatial configuration (e.g., polar outflow and inflow in the disk plane) or dynamical conditions to enable efficient inflow while/after experiencing an outflow.
The kinematically decoupled molecular gas from the galaxy disk just around the sub-mm core (Section~\ref{discussion_keep_nucleus_dusty}) might hint a way of efficient inflow of the molecular gas toward the core by removing the angular momentum.
Some similarities of the NGC~4418 nucleus with the western nucleus of Arp~220 both showing luminous compact core, circumnuclear starburst, and galaxy-scale outflow (Sections~\ref{intro},~\ref{discussion_where_is_starburst}), as well as disturbed circumnuclear kinematics, might suggest that CONs emerge in one phase of the (U)LIRG evolution.

There still remains a possibility of a very young and compact star cluster as an energy source of the core as explored by \cite{sakamoto13} after our optical observational study.
This postulated core cluster is young ($\sim$ a few Myr) enough to be consistent with the nascent starburst scenario proposed for the CON activities (e.g., \citealt{yun01,bressan02,roussel03,rg07}; see Section~\ref{intro}).
Although the superwind-driving starburst is older ($\simeq 10$~Myr old; Section~\ref{consistency_starburst_superwind}) and inconsistent with the nascent starburst scenario, it is found in the circumnuclear region outside of the enshrouded core (Section~\ref{discussion_where_is_starburst}).
Therefore, the nature of the core activity, either a very compact star cluster or an AGN, remains unclear.

\section{CONCLUSIONS}

A nearby luminous infrared galaxy NGC~4418 is known to host a compact obscured nucleus (CON) that shows distinct characteristics such as a very compact sub-mm ($\sim 20$~pc) and MIR core, dusty circumnuclear region with massive molecular gas concentration, flat radio spectral index, and very large FIR-to-radio luminosity ratio.
This galaxy has a ``U''-shaped kpc-scale dust extinct region extending along the disk semiminor axis from the nucleus in the optical color map, and a dusty superwind has been suggested.
We conducted optical spectroscopic study of the nucleus and its surrounding region to understand the nature of the extended gaseous components by analyzing the line ratios and kinematics of various components (ionized gas, cold gas, and stars).
The derived spectroscopic information was used to infer the nuclear activity, starburst, that is responsible for the extended gaseous components.
We then discussed both the starburst and CON activities in the nucleus, as well as the environment of the nucleus that hosts both components.

Shock-excited emission lines are detected up to $\sim 1.7$~kpc from the nucleus toward NW along the disk semiminor axis.
The emission lines there are monotonically blueshifted with increasing distance from the nucleus.
Interstellar Na~D absorption is enhanced around the nucleus and along the NW semiminor axis of the disk.
Its velocity field closely correlates with that of the ionized gas, although their velocity origin at the nucleus shows an offset from the systemic velocity by $\simeq +40$ km~s$^{-1}$.
We concluded that the cold gas is most likely associated with the ionized gas, and that they form a galaxy-scale dusty superwind outflow because the extended nebula in NGC~4418 shows broad similarity to the prototypical superwind in M82 (i.e., dusty conical outflow toward disk polar direction with radially increasing outflow velocity).
In spite of such similarities, they also show some obvious differences.
The superwind in NGC~4418 is more compact in length, less luminous in H$\alpha$ emission, less powerful in outflow velocity and the kinetic energy, and is older in dynamical age than in M82.
It also looks more complex than the one in M82 and shows non-axisymmetric structures in both morphology and velocity field.
The cold-gas outflow rate was estimated to be $\sim 2\ M_\mathrm{\sun}$~yr$^{-1}$ from the cold-gas mass within the superwind ($\simeq 1.8\times 10^7\ M_\mathrm{\sun}$) and dynamical age of the superwind ($\simeq 10$~Myr).
The cold-gas inflow toward the nucleus has been known in earlier absorption line studies, indicating that this galaxy shows both inflow and outflow at the same time.
The inflow rate is roughly comparable to this removal rate, indicating that the supply of the cold and presumably dusty material can keep the dusty nuclear environment.

We also performed spectral synthesis analysis on the SDSS nuclear ($3''$ aperture) spectrum, and found that this nucleus contains a starburst (8--10~Myr old) population (with an additional constraint from the H$\alpha$ luminosity argument).
This starburst is powerful enough to drive the superwind within its dynamical age according to the starburst model based on the {\it starburst99}.
We examined this starburst--superwind scenario and found it consistent with the observations of the nuclear H$\alpha$ emission flux, the infrared luminosity, and the radio continuum fluxes of the extended (500 mas) component.

The sub-mm core and the superwind-driving starburst are not compatible with each other, although both are located within the dusty nuclear region.
If the core is a very compact young star cluster within the core suggested by an earlier sub-mm study for the case of no luminous AGN in the core, its burst age, stellar mass, released energy, and obscuration in the optical wavelengths are not consistent with those of the superwind-driving starburst.
Therefore, the CON nucleus of NGC~4418 is a composite of a compact luminous core and a circumnuclear starburst that has driven the superwind outflow.
Our new observations do not rule our either an AGN or a younger starburst in the core.
Regardless of the true nature of the core, it is difficult to have both core and superwind-driving starburst at the same time when their activities peaked in the same nuclear region.
For the nucleus to have had the starburst $\simeq 10$~Myr ago, the core should not have been luminous at the time of the starburst so that the massive molecular gas can exist in the nucleus to form stars, without being affected by strong UV ionization and/or radiation pressure from the core.
Then the superwind driven by this starburst blew off a significant fraction of the dusty nuclear material.
A subsequent very rapid (in $<10$~Myr) mass inflow toward the very central part of the nucleus is required to generate the dusty, massive, and compact molecular gas concentration ($\sim 10^{8.0}\ M_\mathrm{\sun}$ at $r\lesssim 10$~pc) to bury the luminous core as we see it today.

\acknowledgments

Based on data collected at Subaru Telescope, which is operated by the National Astronomical Observatory of Japan.
Based on data products from observations made with ESO Telescopes at the La Silla Paranal Observatory under programme IDs 177.A-3016, 177.A-3017 and 177.A-3018, and on data products produced by Target/OmegaCEN, INAF-OACN, INAF-OAPD and the KiDS production team, on behalf of the KiDS consortium. OmegaCEN and the KiDS production team acknowledge support by NOVA and NWO-M grants. Members of INAF-OAPD and INAF-OACN also acknowledge the support from the Department of Physics \& Astronomy of the University of Padova, and of the Department of Physics of Univ. Federico II (Naples).
Funding for the SDSS and SDSS-II has been provided by the Alfred P. Sloan Foundation, the Participating Institutions, the National Science Foundation, the U.S. Department of Energy, the National Aeronautics and Space Administration, the Japanese Monbukagakusho, the Max Planck Society, and the Higher Education Funding Council for England. The SDSS Web Site is http://www.sdss.org/.
The SDSS is managed by the Astrophysical Research Consortium for the Participating Institutions. The Participating Institutions are the American Museum of Natural History, Astrophysical Institute Potsdam, University of Basel, University of Cambridge, Case Western Reserve University, University of Chicago, Drexel University, Fermilab, the Institute for Advanced Study, the Japan Participation Group, Johns Hopkins University, the Joint Institute for Nuclear Astrophysics, the Kavli Institute for Particle Astrophysics and Cosmology, the Korean Scientist Group, the Chinese Academy of Sciences (LAMOST), Los Alamos National Laboratory, the Max-Planck-Institute for Astronomy (MPIA), the Max-Planck-Institute for Astrophysics (MPA), New Mexico State University, Ohio State University, University of Pittsburgh, University of Portsmouth, Princeton University, the United States Naval Observatory, and the University of Washington.
The STARLIGHT project is supported by the Brazilian agencies CNPq, CAPES and FAPESP and by the France-Brazil CAPES/Cofecub program.
This work is supported by grant MOST 106-2112-M-001-008- and 107-2119-M-001-026- (Y.O.) and 107-2119-M-001-022- (K.S.).

\facility{Subaru (FOCAS), Sloan} 

\software{STARLIGHT (v04; \citealt{cf05}, E-MILES (v11.0; \citealt{vazdekis16,asad17}}

{}

\end{document}